\documentclass[fleqn,10pt]{wlscirep}
\usepackage[utf8]{inputenc}
\usepackage[T1]{fontenc}

\usepackage{graphicx}%
\usepackage{multirow}%
\usepackage{amsmath,amssymb,amsfonts}%
\usepackage{amsthm}%
\usepackage{mathrsfs}%
\usepackage[title]{appendix}%
\usepackage{xcolor}%
\usepackage{textcomp}%
\usepackage{manyfoot}%
\usepackage{booktabs}%
\usepackage{algorithm}%
\usepackage{algorithmicx}%
\usepackage{algpseudocode}%
\usepackage{listings}%
\usepackage{booktabs}
\usepackage{tabularx}
\usepackage{color}
\usepackage{subcaption}
\usepackage{lineno}
\usepackage{xcolor}
\definecolor{linecolor}{RGB}{139, 38, 53}

\definecolor{mygreen}{rgb}{0.439, 0.678, 0.278}

\makeatletter
\newcommand{\ALOOP}[1]{\ALC@it\algorithmicloop\ #1%
  \begin{ALC@loop}}
\newcommand{\ENDALOOP}{\end{ALC@loop}\ALC@it\algorithmicendloop}

\makeatother

\title{Preventing Unauthorized AI Over-Analysis by Medical Image Adversarial Watermarking}

\author[1,3,+,*]{Xingxing Wei}
\author[2,3,+]{Bangzheng Pu}
\author[1]{Shiji Zhao}
\author[2,3]{Chi Chen}
\author[4,*]{Huazhu Fu}
\affil[1]{Institute of Artificial Intelligence, Beihang University, Beijing, China}
\affil[2]{School of Software, Beihang University, Beijing, China}
\affil[3]{Hangzhou Innovation Institute, Beihang University, Hangzhou, Zhejiang, China}
\affil[4]{Institute of High Performance Computing (IHPC) Agency for Science, Technology and Research (A*STAR), Singapore.}

\affil[*]{corresponding author: xxwei@buaa.edu.cn, fu\_huazhu@ihpc.a-star.edu.sg}

\affil[+]{equal contributions}

\begin{abstract}

The rapid advancement of deep learning has greatly facilitated the integration of Artificial Intelligence  (AI)  into clinical practices, particularly in the realm of computer-aided diagnosis. Given the pivotal role of medical images in various diagnostic procedures, it becomes imperative to ensure the responsible and secure utilization of AI techniques. However, the unauthorized utilization of AI for image analysis raises significant concerns regarding patient privacy and potential infringement on the proprietary rights of data custodians. Consequently, the development of pragmatic and cost-effective strategies that safeguard patient privacy and uphold medical image copyrights emerges as a critical necessity.
In direct response to this pressing demand, we present a pioneering solution named \textbf{M}edical \textbf{I}mage \textbf{Ad}versarial water\textbf{mark}ing  (\textbf{MIAD-MARK}). Our approach introduces watermarks that strategically mislead unauthorized AI diagnostic models, inducing erroneous predictions without compromising the integrity of the visual content within the lesion-free regions. Importantly, our method integrates an authorization protocol tailored for legitimate users, enabling the removal of the MIAD-MARK through encryption-generated keys.
Through an extensive series of experiments, we thoroughly validate the efficacy of MIAD-MARK across three prominent medical image datasets, each corresponding to a significant imaging modality: fundus photography for diabetic retinopathy, microscopic analysis of skin diseases, and magnetic resonance imaging  (MRI)  for brain tumor evaluation. The empirical outcomes demonstrate the substantial impact of our approach, notably reducing the accuracy of standard AI diagnostic models to a mere 8.57\% under white box conditions and 45.83\% in the more challenging black box scenario. Additionally, our solution effectively mitigates unauthorized exploitation of medical images even in the presence of sophisticated watermark removal networks. Notably, those AI diagnosis networks exhibit a meager average accuracy of 38.59\% when applied to images protected by MIAD-MARK, underscoring the robustness of our safeguarding mechanism.

\end{abstract}

\begin{document}

\flushbottom
\maketitle

\newpage

\section*{Introduction}\label{sec1}

Leveraging the considerable surge in computational power, and primarily driven by high-performance GPUs, Artificial Intelligence (AI) coupled with deep learning has achieved substantial advancements and found wide-ranging practical applications. 
These applications have spanned diverse domains~\cite{LeCun2015,Rajpurkar2022,Acosta2022}, encompassing Natural Language Processing (NLP), Computer Vision (CV), robotics, and healthcare. 
Recent progress has witnessed state-of-the-art AI models surpassing their predecessors, capitalizing on foundational model architectures and more expansive datasets, thereby demonstrating enhanced generalization capabilities~\cite{2023arXiv230308774O,Kirillov2023}. However, concerns surrounding the potential misuse of AI-driven data have spurred apprehensions within society, particularly related to data privacy, security, copyright preservation, and the assignment of responsibility~\cite{smith2023stop,ienca2023don,med2023naturewill}. The legal framework governing AI remains in its nascent stages, with governments such as those of the United States, China, and the European Union expressing commitments to safeguard user data privacy and security, necessitating adherence to pertinent legal frameworks by AI developers and providers~\cite{baum2023fear,hutson2023rules}.

In the healthcare domain, AI computer-aided diagnosis (AICAD) holds the promise of elevating medical standards and significantly reducing the costs associated with image-based examinations~\cite{cai2020review, Esteva2019}. 
Conditions requiring specific diagnoses, like dermatological disorders, retinal diseases, and tumors, entail the sharing of sensitive imaging results. However, given the nascent stage of integrating clinical AI systems, uncertainties persist regarding interpretability, accountability, and responsibility\cite{perni2023patients, Price2023}. The unauthorized analysis of medical images introduces concerns over data copyright infringement and breaches of patient privacy\cite{price2019privacy,cacciamani2023artificial}. 
Inadequate protective measures expose these analytical reports to potential access by online service providers, enabling unscrupulous entities to recommend ineffective products or counterfeit medications, thereby leading to financial losses and health complications. Currently, three prominent avenues, blockchain \cite{cheung2021vaccination,guo2021smartphone,szili2023blockchain}, data encryption \cite{yoon2023ehr,yang2022digital}, and digital watermarking\cite{mohanarathinam2020digital}, are being explored for preserving privacy. 
%
%
Among them, digital watermarking is an easy-to-use technique, and hence has wide application in the healthcare domain. Despite that it presents a cost-effective solution for image copyright protection~\cite{Cox2002,939835}, the safeguarding of privacy against unauthorized AI analysis remains a concern. Visible watermarks involve embedding watermark information onto original images using alpha blending, rendering the protected information visible to the human eye and thereby cautioning potential infringers~\cite{liu2010generic,hu2005algorithm,Dekel_2017_CVPR}. On the other hand, invisible watermarks typically employ steganography to embed information for purposes like copyright protection, anti-counterfeiting, and information hiding~\cite{allaf2019review,soualmi2018new,2022An}. 
For medical images, especially those crucial to clinical diagnoses, the preservation of sensitive patient information is paramount. The challenge then becomes: \textit{Is it feasible to simultaneously protect medical image copyright and prevent unauthorized AI models from excessive analysis, thereby achieving cost-effective and proactive privacy protection?}

Recently, adversarial examples, which exploit the vulnerabilities of deep neural networks (DNNs), have been designed to deliberately confuse AI models, leading to erroneous predictions~\cite{Finlayson2019, Zhou2021}. This notion has inspired the exploration of adversarial attacks as a potential avenue to address the issue when AI infringers are involved. Some methods even optimize graphic patterns to generate adversarial examples; for instance, Wei \textit{et al.}~utilize optimized cartoon stickers to deceive human face recognition systems~\cite{wei2022adversarial}, and Jia \textit{et al.}~use logos to generate adversarial watermarks on nature images~\cite{jia2020adv}. However, these methods may neglect the risk of logo removal by infringers and the maintenance of high visual quality for medical professionals.  

In the context of medical image datasets, the challenge is to prevent analysis by unknown models while ensuring the visual quality of lesion regions for medical practitioners. Furthermore, visible watermarks must be robust against removal. This gives rise to three pressing challenges: 
(1) \textbf{Maintaining transferability of the adversarial watermark:} In clinical settings, the details of AI diagnosis models are often unknown, making it challenging to ensure that adversarial watermarks generated for one set of models can also mislead unseen models.
(2) \textbf{Ensuring high visual quality of lesion regions:} Visible watermarking should not compromise key information in the image while still confusing AI diagnostic models. 
(3) \textbf{Preventing unauthorized removal of watermarks:} Visible watermarks are susceptible to removal by advanced Deep Removal Networks (DRNs) \cite{liang2021visible,kim2017splitnet,hertz2019blind,liu2021wdnet}. The challenge lies in developing a method that prevents these networks from successfully removing adversarial watermarks.

\begin{figure}[!t]
\centering
\includegraphics[width=1.0\textwidth]{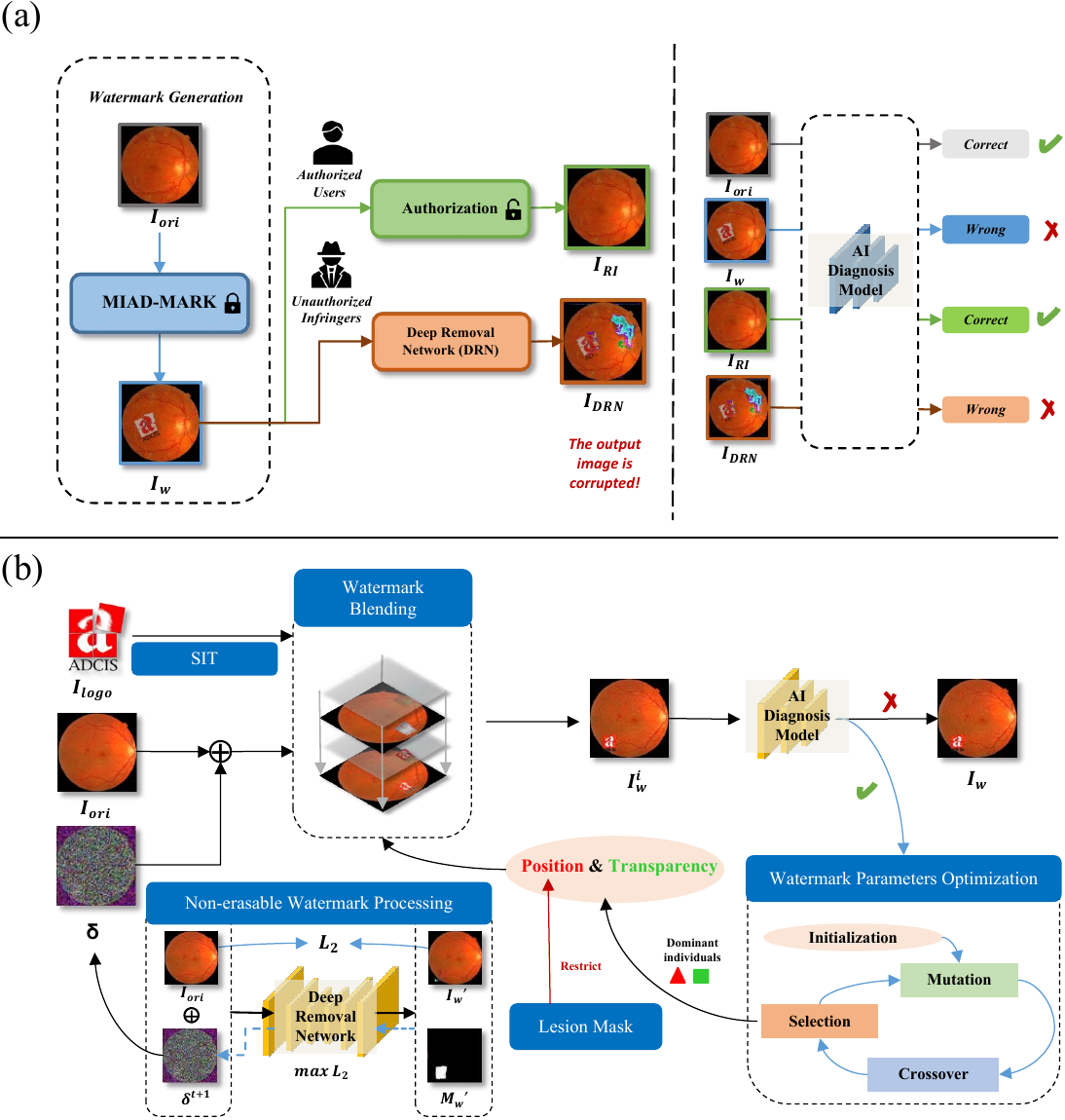} 
\caption{ (a) Overview of our method. The medical image dataset is encrypted by MIAD-MARK, the pipeline to show authorization for authorized users and the measure to prevent infringers from removing watermarks. $I_{ori}$: Original image. $I_w$: Adversarial watermarked image. $I_{RI}$: Watermark removed image. $I_{DRN}$: Deep removal network's output. (b) Framework of MIAD-MARK. SIT: Semantical Invariant Transformation. $\delta$ : Imperceptible perturbation. $I_w^i$: Watermarked image in the $i$-th iteration. $I_w^{'}$: DRN's output in the perturbation optimization process, and $M_w^{'}$ is the predicted mask for watermark region location.}\label{framework}
\end{figure}

To address these challenges, we propose an adversarial visible watermarking technique named \textbf{MIAD-MARK}. This approach aims to protect the copyright of medical images while simultaneously preventing unauthorized AI diagnostic models from excessive analysis. An overview is presented in Figure~\ref{framework}, where MIAD-MARK generates an adversarial watermark image, leveraging its associated watermark logo, to mislead AI diagnostic models and induce false predictions. Additionally, we introduce an authorization process to manage user privileges. Authorized users can remove the adversarial watermark and recover the correctly classified image. Unauthorized infringers may employ advanced DRNs\cite{liang2021visible,kim2017splitnet,hertz2019blind,liu2021wdnet} to erase the logo. Our MIAD-MARK is processed to resist DRN, making its outputs corrupted and unrecognizable to AI models.

\begin{figure}[!t]
  \centering 
  \begin{subfigure}{1.0\textwidth}
    \includegraphics[width=\linewidth]{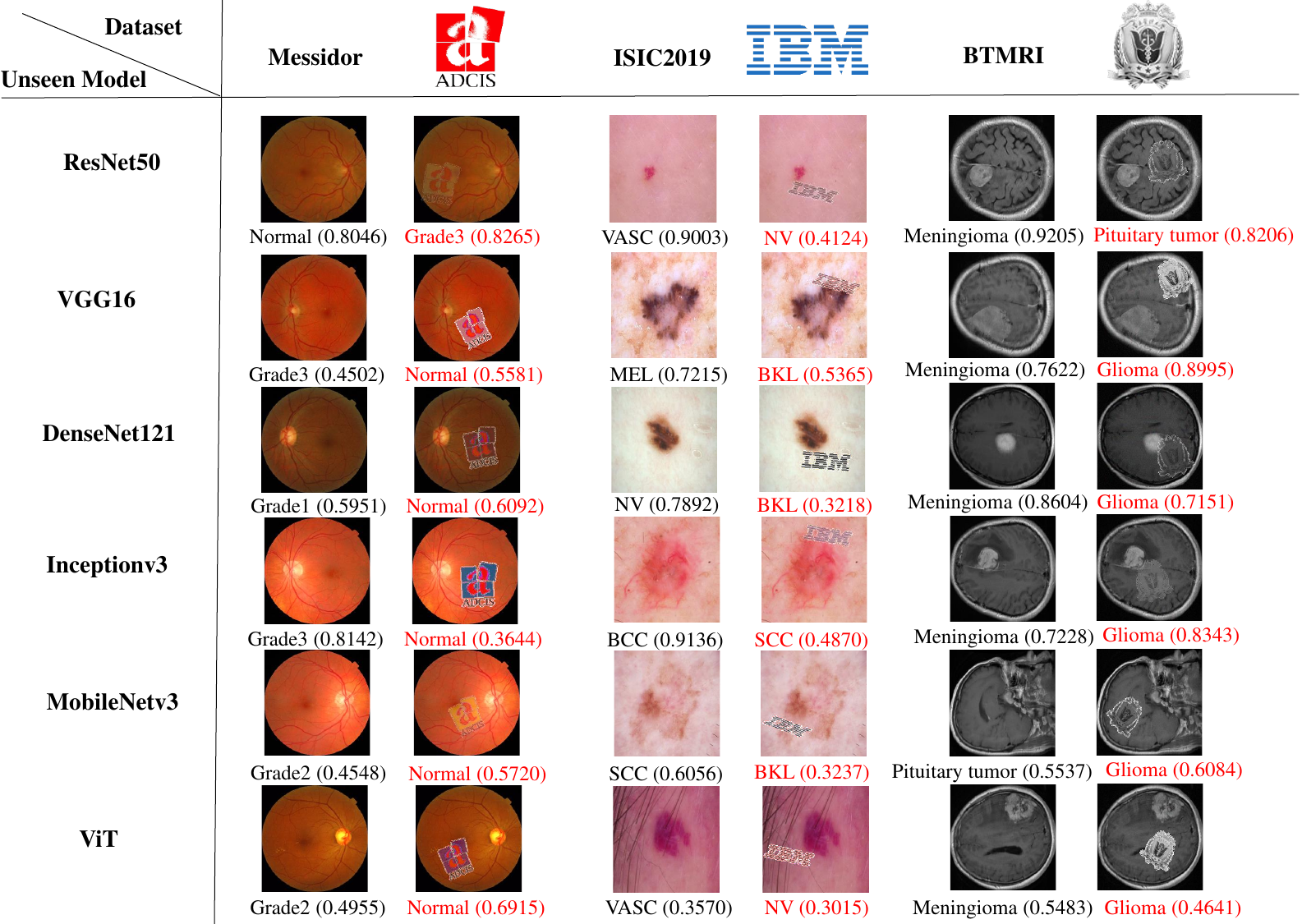} 
    \caption{Visualization of MIAD-MARK on unseen models.}
    \label{fig2a}
  \end{subfigure}
\hfill 
 \begin{subfigure}{0.6\textwidth}
        \centering
        \scalebox{0.65}{
        \begin{tabular}{@{}l|ccc|ccc|ccc@{}}
            \toprule
            & \multicolumn{3}{@{}c@{}}{Messidor} & \multicolumn{3}{@{}c@{}}{ISIC2019}  & \multicolumn{3}{@{}c@{}}{BTMRI}\\
            \cmidrule{2-4}\cmidrule{5-7}\cmidrule{8-10}%
            Seen Model & Random & HE & HE-SIT &  Random & HE & HE-SIT &  Random & HE & HE-SIT \\
            \midrule
            ResNet50  & 0.8103 & \textbf{0.1604} & 0.1656 & 0.9013 & 0.4640 & \textbf{0.2198} & 0.8134	&0.6951	&\textbf{0.6921}
 \\
            VGG16  & 0.7748 & 0.1444 & \textbf{0.0857} & 0.8367 & 0.3451 & \textbf{0.2407} & 0.8894&0.6082	&\textbf{0.4852}
 \\
            DenseNet121 & 0.8731 & 0.3371 & \textbf{0.2491} & 0.8991 & 0.3370 & \textbf{0.1506} & 0.7804 &0.4464 &\textbf{0.3973}
 \\
            Inceptionv3  & 0.7766 & \textbf{0.1351} & 0.1751 & 0.8147 & 0.4454 & \textbf{0.2522} & 0.7759	&\textbf{0.4024}	&0.4623
 \\
            MobileNetv3  & 0.8138 & \textbf{0.2213} & 0.2866 & 0.8669 & 0.4676 & \textbf{0.3208} & 0.9188	&\textbf{0.5583}	&0.5804
 \\
            ViT & 0.8249 & 0.2892 & \textbf{0.1656} & 0.8774 & 0.5079 & \textbf{0.3870} & 0.9261	&\textbf{0.3875}	&0.4122
 \\ \midrule
            Average & 0.8123 &  0.2146& 0.1880&  0.8660& 0.4278& 0.2619& 0.8507& 0.5163& 0.5049
 \\
            Std & 0.0330& 0.0762& 0.0647& 0.0315& 0.0642& 0.0751& 0.0629& 0.1130& 0.1026
 \\
            \bottomrule
        \end{tabular}}
        \caption{MIAD-MARK Accuracy on seen models.}
        \label{tab1}
    \end{subfigure}%
    \begin{subfigure}{0.35\textwidth}
        \centering
        \scalebox{0.65}{
        \begin{tabular}{@{}>{\arraybackslash}m{2.5cm}lcc@{}}
            \toprule
            Unseen Model & Messidor & ISIC2019 & BTMRI \\
            \midrule
            ResNet50     & 0.5360 & 0.4805 & 0.6317 \\
            VGG16        & 0.5478 & 0.6242 & 0.7456
 \\
            DensenNet121 & 0.6726 & 0.5205 & 0.6972
 \\
            Inceptionv3  & 0.4850 & 0.4583 &0.6962
 \\
            MobileNetv3  & 0.5819 & 0.6687 & 0.6978
 \\
            ViT          & 0.5807 & 0.7818 & 0.7566
 \\
            \midrule
            Average & 0.5673 & 0.5890 & 0.7042 \\
            Std & 0.0626 & 0.1251 & 0.0444 \\
            \bottomrule
        \end{tabular}}
        \caption{MIAD-MARK Accuracy on unseen models.}
        \label{ensemble}   
    \end{subfigure} 
  \caption{Results on Seen and Unseen Models.  (a)  The adversarial watermarked images are generated on the ensemble model and attack an unseen model. In each row of images, the original image is on the left side and the adversarial watermark image is on the right side. The predicted class and probability are below each image. \textbf{Black} labels mean the ground truth and \textbf{\textcolor{red}{red}} labels mean wrong predictions. The logo is related to the dataset public logos  ('ADSIC' on Messidor, 'IBM' on ISIC2019, and 'SMU' on BTMRI).  (b)  The performance  (Accuracy)  on seen models for the generated images with different types of visible watermarks, including Random, HE, and HE-SIT. Random: Random parameter watermark, also known as the traditional visible watermark. HE: The watermark is generated by Heuristic Evolutionary. HE-SIT: The watermark is generated by Heuristic Evolutionary with Semantic Invariant Transformation. Std: Standard deviation. (c)  Performance  (Accuracy)  on the leave-one-out unseen model and on HE-SIT settings. Leave-one-out: the unseen model does not appear in ensemble models. A lower accuracy means the adversarial watermarks are more effective against AI models. The original image accuracy for each table is $1.0$.}
  \label{fig:mixed-example}
\end{figure}

\section*{Results}\label{sec2}
\subsection*{MIAD-MARK to protect Medical Images}

In this section, we evaluate MIAD-MARK's performance on seen and unseen models. The seen model setting means the adversarial watermarked images are generated on a known model, and tested on the same model. The unseen model setting means generating adversarial watermarked images on some known models and testing on another unknown model. 

\begin{figure}[!t]
\centering
\includegraphics[width=1.0\textwidth]{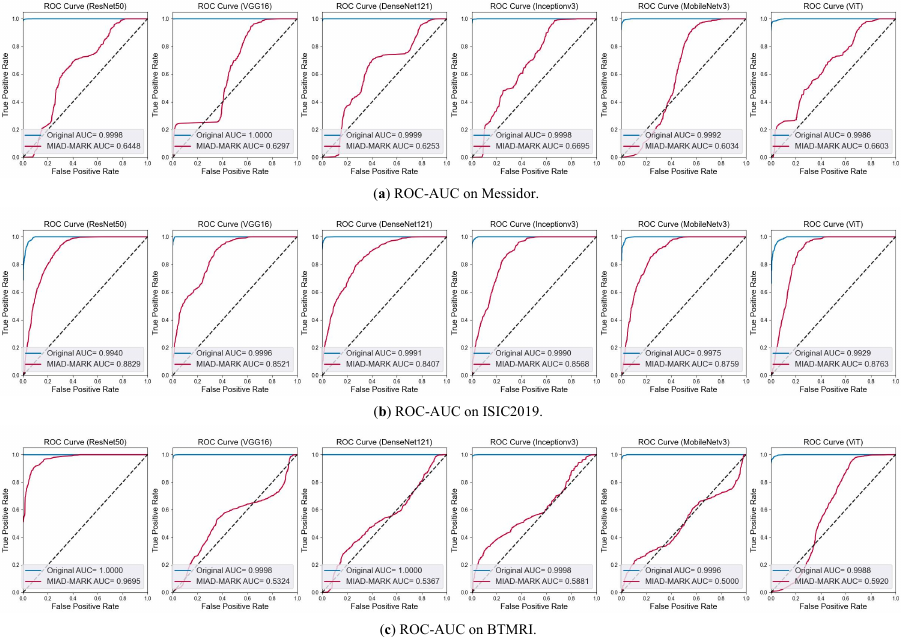}
  \caption{ROC curves of the MIAD-MARKed images on different models and datasets. In each subfigure, the blue line represents the ROC of the original images, and the red line represents the ROC of the watermarking images. The black dotted line is AUC = 0.5, which indicates that the model's performance is equivalent to random guessing. This experiment follows the seen model HE-SIT settings.}
  \label{roc-auc}
\end{figure}

Table~\ref{tab1} shows the quantitative performance of seen AI diagnostic models. 'Random' means watermarks with random positions and transparencies, also called traditional visible watermarks. This watermark can be shown to clarify copyright but can not effectively mislead AI models, the average accuracy(AA) is maintained at 84.30\%. Compared to the 'Random', the Heuristic Evolutionary (HE) algorithm decreases AA to 38.62\%, indicating that it finds the vulnerable watermark parameters for AI models. Additionally, our result shows that HE with semantically invariant transformation (HE-SIT) enhances adversarial performance (AA) to 31.82\%. Thus in other experiments, we choose HE-SIT to optimize the MIAD-MARK's positions and transparencies. SIT, including rotation and HSV (Hue, Saturation, and Value), does not change the semantic information of the logo, so the adversarial watermark can still be easily recognized by human eyes after transformation. As seen in Table~\ref{tab1}, our method achieves the lowest 8.57\% minimum accuracy on Messidor and 15.06\% on ISIC2019 datasets. Accuracy on BTMRI models drops less than the above two datasets, because the lack of color information in grayscale images makes brain tumor MRI more challenging to attack, and the category of BTMRI (three) is less than Messidor (four) and ISIC2019 (eight). Because of category imbalance, we evaluate classifier performance with a more comprehensive metric, ROC-AUC. As seen in Figure~\ref{roc-auc}, the average AUC calculated by all models is 0.6198 on BTMRI, which is slightly lower than Messidor (0.6388) and ISIC2019 (0.8641). This means that adversarial watermarking achieves higher confidence in false categories on fewer-categories classifiers. Despite that, adversarial watermarks (HE and HE-SIT) still significantly decrease the classification accuracy compared to traditional watermarks  (Random).  Our experiments on different backbone DNNs demonstrate that medical image diagnostic models are vulnerable to adversarial watermarks.

In actual application, the AI diagnostic model information is unknown. Therefore, we let the adversarial watermark be created on a source model and then tested on other unknown models. Adversarial watermarked images generated on a seen model perform badly on unseen models. Watermarks produced from a specific seen model tend to overfit their unique characteristics, limiting their transferability. To solve this problem, we employ an ensemble model as the seen model and HE-SIT to generate the watermark. The unseen model is chosen from six AI diagnosis models, each model serves as the unseen model for testing transferability, and the other five models as the ensemble model. As illustrated in Figure~\ref{fig2a}, those adversarial watermarks mislead the AI model's prediction. Table~\ref{ensemble} shows that the ensemble approach drops the average accuracy to 56.73\% on Messidor and 58.90\% on ISIC2019. Nearly half of the adversarial watermarked images maintain the transferability on unseen models, which indicates that generating adversarial watermarks on ensemble models mitigates individual model overfitting. When a watermark generated on a CNN (Convolutional Neural Network) is transferred to a ViT (Vision Transformer), its performance is relatively weak. This is primarily due to the structural differences between CNN and ViT. CNN is sensitive to local features, while ViT excels at capturing global features, which also explains why ViT exhibits stronger robustness than most CNNs on these adversarial watermarks.

\subsection*{Authorization of MIAD-MARKed Medical Images}

\subsubsection*{Authorized Users}
\begin{figure}[!t]
\centering
\includegraphics[width=1.0\textwidth]{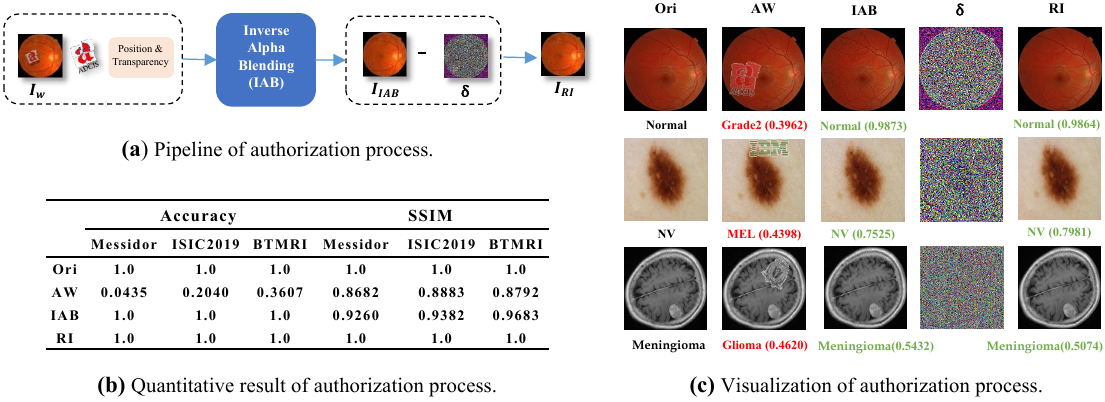}
  \caption{ (a)  The authorization process contains IAB and $\delta$ subtraction, the watermark and  $\delta$  will be removed after this step, and then the clean sample $I_{RI}$ is provided for authorized users.  (b)  The evaluation is to prove if the authorization process can reverse the MIAD-MARK for users. Accuracy is used to metric whether the image can be recognized by AI models  (A higher accuracy means an image close to its original image), and SSIM is used to evaluate the similarity between the original image and others  ($SSIM \in [0,1]$, SSIM close to 1 represents more similar with the original image). Ori: Original images, AW: Adversarial Watermark images, IAB: Inverse Alpha Blending images, $\delta$: perturbation, RI: Recovered Images.  (c)  Visualization of the authorization process, including adversarial watermarks, recovered images, and $\delta$. \textbf{{Black}} labels mean the ground truth,  \textbf{\textcolor{red}{red}} highlights wrong predictions and \textbf{\textcolor{mygreen}{green}} highlights right predictions. }
  \label{authorization}
\end{figure}

For authorized users, we offer a reversible process for MIAD-MARK. In the course of data encryption, we capture a key that encompasses details about watermark placements, transparencies, SIT parameters (if applicable), and the $\delta$ for each image. This enables users to completely remove the watermark logos and $\delta$. Specifically, as shown in pipeline Figure~\ref{authorization} (a), we employ Inverse Alpha Blending (IAB)  to remove the adversarial watermark logos and then minus the $\delta$ to get the recovered clean samples. In this part, our experiments follow the common settings on Appedix~\ref{secA1}, Figure~\ref{figa1}.

As seen in Figure~\ref{authorization} (c) , through authorized
watermark removal process the watermark logos are removed, and the wrong predicted classes are rectified. The recovered images do not include the $\delta$, and their predicted classes are the same as the ground truth labels. To quantitatively compare recovered images with original images, we use Accuracy and SSIM to evaluate recovered images on AI models and visual differences. In Table~ \ref{authorization} (b), the original images' $Accuracy$ is 1.0, the adversarial watermarks' $Accuracy$ is 0.0, and the recovered images' $Accuracy$ is 1.0. This indicates that recovered images can be correctly recognized after our authorization. The recovered images' SSIM is equal to the original images, indicating that the recovered images are the same as the original images on the structure. Adversarial watermark images' SSIM decreased because of $\delta$ and watermark logos. However, SSIM still exceeds 0.8, which means MIAD-MARK has a low impact on vision, so it meets the requirement of maintaining a high visual quality of visible watermarks.

\subsubsection*{Unauthorized Users}

\begin{figure}[!t]
  \centering 
  \begin{subfigure}{0.8\textwidth}
    \includegraphics[width=1\textwidth]{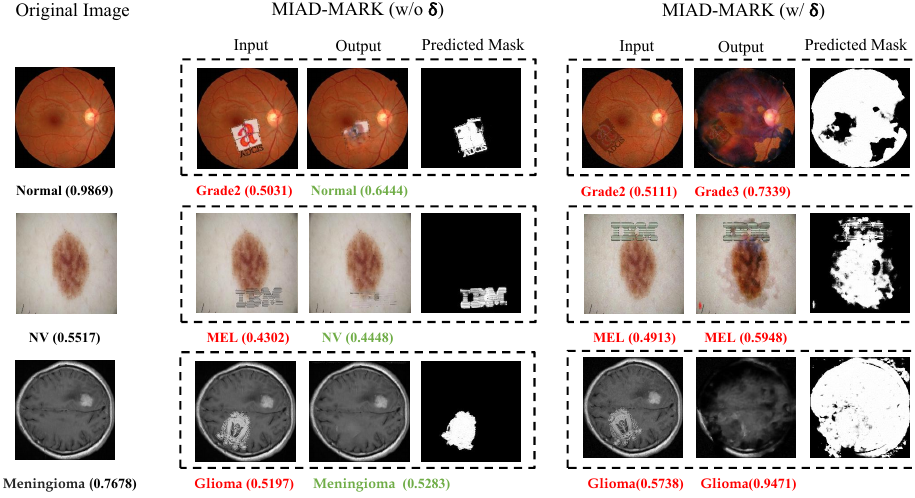} 
    \caption{Visualization of adversarial watermarks through the deep removal network (DRN).}\label{wv}
  \end{subfigure}
  \hfill 
   \begin{subfigure}{1.0\textwidth}
        \centering
        \scalebox{0.8}{
       \begin{tabular*}{\textwidth}{@{\extracolsep\fill}lccccccc}
          \toprule%
          & & \multicolumn{2}{@{}c@{}}{Messidor} & \multicolumn{2}{@{}c@{}}{ISIC2019}  & \multicolumn{2}{@{}c@{}}{BTMRI}\\\cmidrule{3-4}\cmidrule{5-6}\cmidrule{7-8}%
          Metric & Watermark type& SplitNet & WDNet &  SplitNet & WDNet &  SplitNet & WDNet\\
          \midrule
          \multirow{2}*{SSIM $\downarrow$}&w/o $\delta$ &0.9706 & 0.9698&0.9698&0.9766& 0.9842&0.9839 \\
          & w/ $\delta$ &\textbf{0.7110} &\textbf{0.3579}&\textbf{0.7666}&\textbf{0.7732}&\textbf{0.7732}& \textbf{0.4705}\\
          \midrule
          \multirow{2}*{PSNR $\downarrow$}& w/o $\delta$ &44.6426 &43.8920 & 45.9274 &45.7423 &47.9332 &47.8375 \\
          & w/ $\delta$ &\textbf{33.4627}	&\textbf{29.6201} &\textbf{33.4859}	&\textbf{29.1537} &\textbf{32.9987} &\textbf{29.5853} \\
          \midrule
          \multirow{2}*{RMSE $\uparrow$} &w/o $\delta$ &1.6108 & 1.6954 &1.3066 &1.3632 &1.0275&1.0378 \\
          & w/ $\delta$ &\textbf{5.4253} & \textbf{8.5274}&\textbf{5.4092}&\textbf{8.9949}&\textbf{5.7385}&\textbf{8.5061}\\
          \midrule
        \end{tabular*}}
 
        \caption{Quantitative evaluation of the visual quality.} \label{quality}         
    \end{subfigure}    
    \hfill
 
    \begin{subfigure}{1.0\textwidth}
        \centering
        \scalebox{0.8}{
            \begin{tabular*}{\textwidth}{@{\extracolsep\fill}lccccccc}
                \toprule%
                & & \multicolumn{2}{@{}c@{}}{Messidor} & \multicolumn{2}{@{}c@{}}{ISIC2019}  & \multicolumn{2}{@{}c@{}}{BTMRI}\\\cmidrule{3-4}\cmidrule{5-6}\cmidrule{7-8}%
                Model& Watermark type & SplitNet & WDNet & SplitNet & WDNet & SplitNet & WDNet\\
                \midrule
                \multirow{2}*{ResNet50} &w/o $\delta$ & 0.3029 & 0.3143 & 0.7896 & 0.6759 & 0.8460 & 0.8553 \\
                &w/ $\delta$& \textbf{0.2875} & \textbf{0.2944} & \textbf{0.2352} & \textbf{0.2609} & \textbf{0.7770} & \textbf{0.3846} \\
                \midrule
                \multirow{2}*{VGG16}& w/o $\delta$& 0.4011 & 0.3531 & 0.6375 & 0.4286 & 0.7873 & 0.8120 \\
                &w/ $\delta$ & \textbf{0.2710} & \textbf{0.2290} & \textbf{0.3212} & \textbf{0.2585} & \textbf{0.3864} & \textbf{0.6595} \\
                \midrule
                \multirow{2}*{DenseNet121}&w/o $\delta$ & 0.4800 & 0.6320 & 0.6851 & 0.5776 & 0.8140 & 0.8813 \\
                &w/ $\delta$ & \textbf{0.3741} & \textbf{0.5327} & \textbf{0.2275} & \textbf{0.1908} & \textbf{0.1966} & \textbf{0.6595} \\
                \midrule
                \multirow{2}*{Inceptionv3}&w/o $\delta$ & 0.4903 & 0.5440 & 0.7496 & 0.6467 & 0.7080 & 0.7440 \\
                &w/ $\delta$ & \textbf{0.3755} & \textbf{0.3333} & \textbf{0.3059} & \textbf{0.2319} & \textbf{0.4800} & \textbf{0.2963} \\
                \midrule
                \multirow{2}*{MobileNetv3}&w/o $\delta$ & 0.4320 & 0.5097 & 0.6037 & 0.5284 & 0.6513 & 0.7846 \\
                &w/ $\delta$ & \textbf{0.2999} & \textbf{0.3832} & \textbf{0.2868} & \textbf{0.2367} & \textbf{0.6034} & \textbf{0.6510} \\
                \midrule
                \multirow{2}*{ViT}&w/o $\delta$& 0.5040 & 0.5691 & 0.5806 & 0.5238 & 0.7620 & 0.7420 \\
                &w/ $\delta$ & \textbf{0.4594} & \textbf{0.4829} & \textbf{0.4015} & \textbf{0.2609} & \textbf{0.6391} & \textbf{0.5912} \\
                \midrule
            \end{tabular*}}
        \caption{Accuracy on the DRN's outputs.}\label{acc1}        
    \end{subfigure}    

    \caption{ (a) Visualization of adversarial watermarks through the Deep Removal Network (DRN). $w/ \delta$: adding perturbation to protect MIAD-MARK, w/o $\delta$: MIAD-MARK without adding perturabtion. Input: inputs of DRN, Output: outputs of DRN. The predicted mask shows the result of the watermark region network in DRN. \textbf{{Black}} labels mean the ground truth,  \textbf{\textcolor{red}{red}} highlights wrong predictions and \textbf{\textcolor{mygreen}{green}} highlights right predictions. (b)  Quantitative evaluation of the visual quality. SplitNet and WDNet are two types of DRNs. (c) Accuracy of DRN's outputs. The bold digital is the lowest accuracy in each model.}
    \label{unautho}
\end{figure}

For unauthorized users, the images protected by MIAD-MARK are unavailable for AI models. We assume that data infringers attempt to find a way to erase adversarial watermarks and restore images. However, manually removing watermarks with image processing software is impractical. Deep Removal Networks (DRNs)\cite{liang2021visible,kim2017splitnet,hertz2019blind,liu2021wdnet} can be used to remove watermarks with high efficiency, thus showing a great threat to MIAD-MARK. In this part, we conduct experiments to verify that adversarial imperceptible perturbation, which is an effective approach to prevent adversarial watermarks removed by DRNs.

Figure~\ref{wv} shows the impact of $\delta$  on the outputs of DRN. For human eyes, the $\delta$ of images is invisible. $\delta$ is an imperceptible perturbation, which is produced by gradient-based optimization. In order to eliminate the negative impact on the adversarial watermark logo, $\delta$ is added to the original image before the adversarial watermark generation, see detail on Appendix~\ref{secB}. As seen in Figure~\ref{wv}, watermark logos are erased on the ‘w/o $\delta$’ column, and the misclassified labels are rectified. On the contrary, the outputs of the ‘w/ $\delta$’ are corrupted and misclassified. As shown in the predicted masks, watermark regions of the ‘w/o $\delta$’ inputs are correctly located while the ‘w/ $\delta$’ failed. Their masks are larger and cover the lesion region, so the refined net of DRN process these unwanted part and disrupt the image.

Table~\ref{quality} evaluates the visual qualities between DRNs' outputs and original images. $SSIM$ on the ‘w/o $\delta$’ row is close to 1, which indicates DRN can effectively remove unprotected watermarks. However, $SSIM$ on the 'w/ $\delta$' row decrease significantly indicates the outputs of DRN have been changed obviously compared to the original images. $PSNR$ on the ‘w/ $\delta$’ row is lower than the ‘w/o $\delta$’, meaning that $\delta$ causes significant differences in the output of DRN. $RMSE$ of ‘w/ $\delta$’ is several larger times than ‘w/o $\delta$’ $RMSE$, indicating that outputs of DRN are modified severely, not only a small area of the watermark logo. 

Table~\ref{acc1} shows the AI diagnosis models' predictions after the DRNs' process. The accuracy of the ‘w/ $\delta$’ is lower than the ‘w/o $\delta$’, which indicates $\delta$ has a negative impact on the DRN's performance.
In consideration of visualization results and descend accuracy, the $\delta$ can effectively protect our watermarked images. Even if the infringers attempt to erase our watermark logos, the output images are unanalyzable for humans and AI.

\section*{Discussion}\label{discussion}
The latest advancement in the field of digital healthcare is AI-assisted diagnosis, benefiting from recent progress in large-scale models. PMC-llama \cite{wu2023pmcllama} and Med-PALM \cite{singhal2023large} have demonstrated clinical expert-level capabilities in diagnosing diseases. In online AI diagnostics, patients are required to upload imaging results to their electronic medical records. Such services have significantly increased the exposure risk to patient privacy. In the past few years, privacy breach incidents in non-clinical scenarios have raised wide concerns. Technology corporations employ AI-based data analysis techniques to analyze vast user datasets, thereby influencing company decisions or reaping profits. Examples include personalized content and advertisement delivery to users, as well as the development of new products. Despite many companies claiming to implement protective measures for user privacy, the risk of privacy infringement still persists. Instances of unauthorized AI model analysis of user privacy have led to severe consequences. Cambridge Analytica, for instance, provided services to multiple political campaigns by illicitly acquiring personal data from approximately 87 million Facebook users through a psychological test application. Leveraging this data alongside their developed AI models, the company delivered tailored political advertisements to users in an attempt to influence their voting behaviors \cite{rosenberg2018, cadwalladr2018}. Therefore, it is essential to develop technologies that protect patient privacy against these new challenges. Furthermore, some patient data has been knowingly shared and authorized by patients for use in clinical research or training AI models. The demand for protecting the copyright of these datasets has also emerged.

Hence, our research endeavors to develop a cost-effective digital watermarking solution to protect image copyrights and patient privacy. Additionally, it involves related authorization and verification processes. To achieve AI model misdirection while safeguarding image copyrights, we leverage the concept of adversarial examples and incorporate logo watermarking to create adversarial watermarks. In the white-box setting (seen models), MIAD-MARK performs more effectively compared to the black-box setting (unseen models). This observation aligns with the prevalent outcomes observed in most prior adversarial attack studies\cite{akhtar2018threat, zhang2019adversarial}. MIAD-MARK generated from seen models can adapt to changes or updates in the target model by dynamically adjusting the strategy for generating adversarial samples through real-time access to the model's latest information. In contrast, MIAD-MARK attack unseen models can only rely on the previously collected seen model information to generate adversarial samples, which elevates the risk of adversarial example obsolescence. Given our focus on online service diagnostic models, where the specific model information is unavailable. We improve the MIAD-MARK based on ensemble models, and the result shows a significant improvement in transferability compared with the single model performance. This phenomenon is akin to ensemble reasoning, wherein during the generation of adversarial watermarks, the utilization of ensemble models can enhance generalization capability, mitigate the risks of overfitting or underfitting, and balance the biases and variances inherent in different foundational models \cite{liu2019deep}.

Building upon adversarial effectiveness, we consider that MIAD-MARK must also possess visibility without disturbing the original image and exhibit resistance against erasure.
Mintzer et. al. \cite{mintzer1997effective} point out that visible watermarking should be clearly visible to human eyes and not obviously block important objects in the image. Thus our watermark obeys the rule that does not hinder medical professionals' diagnosis of images and must avoid obscuring pathological findings. We devise the GradCAM-guided methods to ensure watermarks avoid crucial regions, GradCAM offers an intuitive way for humans to understand the attention of CNN for specific classes \cite{selvaraju2016grad}. In the medical image domain, gradient-weighted feature maps have been shown to provide initial localization of lesions \cite{chien2022usefulness,he2019medimlp,panwar2020deep}. We employed a threshold to control the size of the constrained region, and this threshold significantly impacts MIAD-MARK's performance. A larger constrained region can result in the inability to effectively search for the adversarial watermark area. Conversely, a too-small constraint would undermine the principle of maintaining the visual quality of the important regions on original images. Due to numerous images in datasets and medical domain knowledge, it is time-consuming to manually erase the watermarks clearly, even if assisted by semi-automatic photograph processing software. 

However, visible watermarks are at actual risk of removal by automatic image restoration techniques. Early image processing approaches such as inpainting \cite{huang2004attacking} or image matching \cite{park2012identigram}, are less effective in watermark localization and image restoration, the recovered image exists a visual gap compared with the original. Recently, two-stage methods based on DNN narrowed this gap. Firstly, DRNs predict rough decomposition of the watermarked region and then refine the removal results\cite{liang2021visible,kim2017splitnet,hertz2019blind,liu2021wdnet}. Thus it is imperative to enhance watermark robustness to prevent deep removal. Adversarial perturbations can be conducted to interrupt such removal processes. They are optimized by Projected Gradient Descent \cite{madry2017towards}  (PGD), and the loss function measures the $L_2$ distance between DRNs' output and the original images, and maximizing this function will lead to a large difference between these two images. Perturbation magnitudes are restricted by the infinite norm to be imperceptible. For authorized users, MIAD-MARK is reversible, they can get clean samples by Inverse Alpha Blending  (IAB)  with a key and straightly subtracting $\delta$. This key includes alpha blending parameters for each watermarked image. For unauthorized users, medical images with adversarial watermarks are misclassified by AI models. However, infringers can exploit advanced two-stage DRNs to remove watermarks and recover the image. By observing the predicted regions of the watermark logo\ref{wv}, $\delta$ can make DRNs locate untargeted regions and produce wrong alpha blending compositions, resulting in image corruption. Thus, MIAD-MARK serves as a robust adversarial watermark that displays copyright information without compromising doctors' visual assessment and preserves patient privacy against AI over-analysis.

Our approach comes with certain inherent limitations that merit consideration. \textbf{Firstly}, the unavailability of open-access online medical image diagnostic platforms necessitated our utilization of widely recognized models with distinct architectures as the backbones for AI diagnostic models. These models stand as representative examples in their respective application domains. For instance, ResNet is prevalent in CNN-based diagnostic frameworks \cite{wang2020detection,kim2019deep,isensee2018automated}, DenseNet's dense residual connections enable deeper CNN networks, yielding superior results on larger medical datasets such as chest X-rays \cite{9478715}, and MobileNet serves as a lightweight model tailored for edge devices \cite{abid2021optimizing}. Meanwhile, ViT's prominence among larger models is highlighted by its extensive parameter count and consequent efficacy in extensive data training \cite{oquab2023dinov2}. While our ViT model is smaller than its larger counterparts, its fundamental components like multi-head attention mechanisms remain indicative of the architecture. Nonetheless, it's important to acknowledge that these chosen model architectures may differ from those employed in actual real-world scenarios, underscoring the necessity for ongoing evaluations involving real-world models in the future.
\textbf{Secondly}, it is noteworthy that ViT proves to be more resilient against attacks compared to most CNNs. This observation indicates that large-scale transformer-based vision models demonstrate heightened robustness in the face of MIAD-MARK. This perspective is substantiated by prior adversarial attack studies on histopathological classification models \cite{ghaffari2022adversarial}. CNN's local receptive field renders it more susceptible to attacks like MIAD-MARK, which focuses on specific image regions. Conversely, ViT relies on a self-attention mechanism for establishing long-range dependencies and utilizes positional encoding to encapsulate the global spatial context of the image \cite{dosovitskiy2021image}, rendering it more robust against adversarial examples. The architectural disparities between these models contribute to the complexities associated with achieving robust attacks.

\section*{Conclusion}\label{sec12}

In this paper, we introduce MIAD-MARK, a visible watermarking technique designed to safeguard the copyright of medical datasets and uphold patient privacy. MIAD-MARK achieves this dual purpose by providing clear ownership identification warnings to potential infringers, all the while preserving the diagnostic capabilities of unauthorized medical image analysis models. Moreover, we demonstrate the resilience of MIAD-MARK against advanced deep watermark erasure techniques through the strategic introduction of perturbations.
In summary, our work addresses a pivotal intersection of AI, medical imaging, and privacy concerns. By introducing the MIAD-MARK technique, we contribute a pragmatic and effective approach to mitigate the inadvertent risks associated with AI-driven image analysis. In doing so, we endeavor to foster a more secure and ethically responsible landscape for the application of clinical AI, thereby enhancing the integrity of this critical field.

\section*{Methods}\label{sec11}

The pipeline of the MIAD-MARK is shown in Figure~\ref{framework}b. Firstly, we prepare the medical image and its logo  (illustrated as the fundus photograph and ADCIS). The original image undergoes a preprocessing step where the perturbation $\delta$ is introduced, establishing a non-erasable characteristic. The logo is conducted to semantic invariant transformations (SIT). Subsequently, the watermark and the original image are blended, based on position and transparency, to create the input for the DNN-based AI diagnosis model. The watermark's placement is constrained through the introduction of a lesion mask. The dominant position and transparency are determined through a watermark parameter optimization process. A wrong prediction signifies the successful generation of an adversarial watermark. If DNN makes a correct prediction, the optimization process persists until produces an effective adversarial watermark or the designated loop ends.

\subsection*{Non-erasable Watermark Processing} \label{sec35}
Inspired by adversarial attacks, Liu et al. \cite{liu2022watermark} propose to produce artificially designed perturbations against DRNs. They update the perturbation during gradient back-propagation and add it to protect the original image. The perturbation is calculated by projected gradient descent (PGD) \cite{madry2017towards}, which can be formulated as,
\begin{equation}             
     \delta^{t+1} = Proj (\delta^{t}+\sigma \textbf{sign} (\bigtriangledown \mathcal{L}  (I_w,\delta^t) ) ) ,
        \label{eq1}
\end{equation}
where $\delta^{t}$ means the perturbation in $t$-th iteration. The sign is the direction of the gradient and the value of $\delta^{t}$ is restricted by the $L_{\infty}$ norm bound $\varepsilon$. $\sigma$ is the updated step.

To ensure the removal network can successfully erase watermarks and recover the image on w/o$\delta$ images, we freeze the encoders and fine-tune decoders with a training set. This process involves minimizing the mean squared error loss between the original image and the output of the watermark removal network:
\begin{equation}             
    \min {\mathcal{L}_{MSE}}=\frac{1}{mn}\sum_{i=0}^{n-1}\sum_{j=0}^{m-1} (I_{w^{'}}-I_{ori}) ^2,
        \label{eq2}
\end{equation}
where $I_{w^{'}}$ represent the output of watermark removal network. $n,m$ represents image width and height. 

By adjusting the optimization objectives, $\delta$ can disrupt outputs or immune erasure. Thus there are two types of $\delta$: Perturbation of Disrupting the images $\delta_D$ and Perturbation of Inerasable Watermark $\delta_I$. In the main context of this paper, we adopt the $\delta_D$ because it is more efficient for protecting the watermark logo, the detailed analysis can be seen in Appendix~\ref{secB}. And in the result and discussion of this paper, we use  $\delta$  to simply represent $\delta_D$. The optimization objective of $\delta_D$ is to maximize the difference between the original image and the output, leading to output corruption. This process can be formulated as,
\begin{equation}             
    \max \mathcal{L}_{\delta_D} = \lVert I_{ori} - I_{w'} \rVert_2.
    \label{eq3}
\end{equation}
The perturbation$ \delta$ is updated by Equation~\ref{eq1} in each step until the max iteration.

$\delta_I$, on the other hand, minimizes the difference between the original image and the output while using a regularization term to reduce the distance between the mask of the predicted watermark and the all-zero mask, thus making the watermark indelible. This process can be expressed as,

\begin{equation}             
    \min \mathcal{L}_{\delta_I} = \frac{1}{2}  (\gamma \lVert I_{ori} - I_{w'} \rVert_2 +  \lVert M_{w'} - M_{0} \rVert_2) ,
    \label{eq4}
\end{equation}
where $\gamma$ is used to balance the two terms of $\mathcal{L}_{\delta_I}$. $M_{w'}$ is the predicted mask of watermark, and $M_{0}$ is a zero mask.

\subsection*{Watermark Blending}
The alpha blending technique \cite{shen1998dct} is used to create the translucent effect by compositing a foreground with a background. Thus we can blend the watermark logo with an image to claim the copyright. As seen in Algorithm~\ref{algo1}, given the medical image $ (I_{ori}^{W\times H\times C}+\delta) $ and a watermark logo ${I_{logo}}^{w\times h\times C}$, where $W,w$ are wide, $H,h$ are height, and $C$ is channel, including $RGB$ channel and alpha channel $A$. First, we adjust the size of $I_{logo}$ to ensure an appropriate proportion
between the logo and the background, this can be calculated by:
\begin{equation}
    [w',h']= min (\frac{W}{sl \cdot w},\frac{H}{sl \cdot h})  \times [w,h],
    \label{eq5}
\end{equation}
where $sl$ represents the scaling ratio. Intuitively, taking ${I_{ori}}^{224\times 224}$ and ${I_{logo}}^{441\times 245}$ as an example, when $sl=1$, the logo occupies about $14\%$ of the original image area, when $sl=2$ it occupies $7\%$, and when $sl=4$, it occupies $3.5\%$. We define the the upper-left point of $I_{ori}$ is the original coordinate and upper-left point of $I_{logo}$ is  (x,y) , then generate a foreground ${I_f}^{W\times H}$ by pasting ${I_{logo}}^{w^{'}\times h^{'}}$ on a zero background ${I_b}^{W\times H}$ from $[x,x+w^{'}],[y,y+h^{'}]$. To keep the background of $I_{ori}$ clean, we convert $I_f$ to a binary mask $I_m$. In this way, the watermark image can be calculated as:
\begin{equation}
    I_w=\alpha \cdot I_f +  (1-\alpha) \cdot I_m\cdot  (I_{ori}+\delta) +  (1-I_m) \cdot  (I_{ori}+\delta) ,
    \label{eq6}
\end{equation}
where $\alpha$ represents transparency,$\alpha \in [0,1]$.

\subsection*{Lesion Mask Generation}
The lesions or important anatomical positions are critical to diagnosing diseases for doctors. Semantic segmentation models can generate precise pixel-level masks by learning labels from clinical experts. However, it is hard to collect sufficient pixel-level labels to train segmented models for each disease of medical imaging datasets. Alternatively, we adopt
Gradient-weighted Class Activation Mapping  (Grad-CAM)  to create disease-related masks, which highlight the region of interest for model prediction. Grad-CAM can be easily integrated into any CNN classifier without modification. Thus it supplies a powerful tool to help us limit watermarking positions. 

As seen in Algorithm~\ref{algo2}, Grad-CAM heatmap is activated by weighted features of a certain convolutional layer, which gives the non-attention model localization ability. The importance weight of features can be calculated by logits' gradient of ground-truth label $l_g$ and global average pooling, this can be formulated as,
\begin{equation}
    \omega_n^{l_g}=\frac{1}{N}\sum_i\sum_j\frac{\partial f^{l_g}}{\partial A_{ij}^n},
    \label{eq7}
\end{equation}
where $f$ represents logits of $l_g$ before softmax, $A_{ij}$ is the feature map value in $ (i,j) $, $N=i\times j$. Then we sample features with sampling rate $r_c$ from the target convolutional layer and select channels with high weight ranking. The Grad-CAM heatmap is calculated by the sum of weighted features, this can be expressed as,
\begin{equation}
{I_{Crad-CAM}}=Relu (\sum_k{\omega_k \cdot A_k}) ,
    \label{eq8}
\end{equation}
where Relu is Rectified Linear Unit, $k\leq n$.

Finally, we resize ${I_{Crad-CAM}}$ to $W\times H$ and binary with threshold $t$ to get mask $I_M$, then generate bounding boxes for each connected region to limit the watermark position. 

\subsection*{Watermark Parameters' Optimization}

To prevent unauthorized medical diagnostic model analysis, our core idea is to generate adversarial watermarking to mislead the classifier. According to Algorithm~\ref{algo1}, transparency and watermark location are the two main parameters that influence alpha blending. Thus we optimize these parameters and create adversarial watermarks until the source models produce the wrong prediction. Inspired by natural evolutionary strategies, the evolutionary algorithm \cite{back1993overview} is generally used for discrete search. The traditional evolutionary algorithm includes three main steps:  (1) initializing parental generation;  (2) mutation and crossover to produce abundant offspring;  (3) selecting dominant individuals. 

In this paper, we employ a heuristic evolutionary algorithm to reduce the search space. Inspired by the location aggregation effect of patch attack \cite{wei2022adversarial}, we downsample a series of watermark positions with uniform and sparse distribution. Then the offspring are selected according to their fitness. In order to enrich the diversity of the population, our approach adopts mutation of adaptive factors to generate new genes and retain part of the parental traits through gene hybrid.

In our Algorithm~\ref{algo3}, the smallest unit called the gene, makes up the individual, and the collection of individuals forms a population. Specifically, each individual is a triplet, represented by $\rho=[x,y,\alpha]$. The initial population is called parents $P$.   

The number of parental population $N_p$ is controlled by a point sampling rate $r_p$. The sampled number $ N=\lfloor r_p\cdot (W-w)  (H-h) \rfloor$  ($\lfloor \rfloor$ means round down) , let $N=N_x\cdot N_y, \frac{N_x}{N_y}=\frac{W-w}{H-h}$. The row number and column number are $N_x=\lfloor \sqrt{N\cdot\frac{W-w}{H-h}}\rfloor,N_y=\lfloor\sqrt{N\cdot\frac{H-h}{W-w}}\rfloor$, and the gap between two adjacent point calculated by $d_x=\frac{W-w}{N_x}, d_y=\frac{H-h}{N_y}$. Thus the location of parents can be derived as,

\begin{equation}
     (x_n,y_m) = (n\cdot \lfloor\sqrt{\frac{1}{r_p}}\rfloor, m\cdot \lfloor\sqrt{\frac{1}{r_p}}\rfloor) ,
    \label{eq9}
\end{equation}
where $n=0,...,N_x, m=0,...,N_y, 0<r_p\leq 1$. 

We define the number of populations as $Np$, and the maximum generation is $N_g$. After determining the alternative points of parents using the above method, we select $Np$ dominant individuals as parents based on confidence. Next, each parental individual will generate offspring that is closer to their nearest neighbor in mutation. The mutation step $s_p$ is determined by maximum generation $N_g$ and gap distance $d$,

\begin{equation}
s_p= \left\{
        \begin{aligned}
        &\lfloor\frac{d}{2\cdot N_g}\rfloor, \text{if} \ d>2\cdot N_g \\
        &1, \text{otherwise}.
        \end{aligned}
        \right.
    \label{eq10}
\end{equation}
Transparency moving step represented by $s_\alpha$, thus mutation step $s=[s_{p_x},s_{p_y},s_\alpha]$.
The $k$-th mutated gene of $j$-th individuals in the $i$-th generation expressed as,

\begin{equation}
    \rho_{ij}^z\in \{ \rho_{{ (i-1) }j}^z, ~ \rho_{{ (i-1) }j}^z + s_{ij}^z,~ \rho_{{ (i-1) }j}^z - s_{ij}^z \},
    \label{eq11}
\end{equation}
where $i=1,2,...,N_g$, $z=0,1,2$. Following mutation, a novel gene is produced and crossover with the existing gene in a ratio $r_c$. The maximum number of possibilities for each individual is gene numbers to the third power $j=0,1,...h$, $h<[len (\rho) ]^3 \cdot N_p$, with the population size $k$ controlled by multiplying the crossover ratio, $k<[len (\rho) ]^3 \cdot N_p \cdot r_c$.

To select dominant individuals, the cross-mutated individual is compared with the best individual from the preceding generation and $N_p$ dominants. This process results in the formation of a new population. If no individual is found to cause misclassify, the mutation and crossover procedures are iteratively applied to produce additional offspring until either the model yields an incorrect output or the maximum algebraic stop is reached.

\subsection*{Transferable Adversarial Watermark}
In practice, the information about unauthorized medical image diagnosis systems is unknown, thus posing a challenge to the transferability of the adversarial watermark.
Transferable means that adversarial watermarks generated by querying the source model can also be effective on other models. We select a bunch of mainstream backbone DNNs as the library of source models and evaluate the transferability of watermarking images. To enhance transferability, we ensemble the source model during watermark generation and launch attacks on it. The ensemble confidence of the image $I_n$ can be calculated as,
\begin{equation}             
      f (I_n) _{ensemble}=\sum_{i=1}^{k} \beta_if_i (I_n) 
        \label{eq12},
\end{equation}
 where $k$ is the number of models, and $\beta$ is the weight of single model.

\subsection*{Ownership authorization}
Access to the datasets during deployment is controlled through unique keys, which are specialized documents containing watermarking parameters assigned to authorized users. $Key=\{I_0:[x_0,y_0,\alpha_0], ..., I_n:[x_n,y_n,\alpha_n]\}$. To remove the watermarks, authorized users may employ an inverse alpha blending operation. Incidentally, If the semantically invariant transformation is implemented, the logo is used as transformed. After inverse alpha blending, the watermark logo is erased, then minus the $\delta$ can recover to the original images. This process can be formulated as, 
\begin{equation}             
     I_{ori}=  (I_{w}-\alpha \cdot I_f)  / (1-\alpha \cdot I_m) -\delta.
        \label{eq13}
\end{equation}







\subsection*{Experimental Settings}
\subsubsection*{Datasets}
We conduct experiments on three widely-used and challenging medical imaging datasets: Messidor \cite{decenciere2014feedback} diabetic retinopathy, ISIC2019 \cite{combalia2019bcn20000}, and Brain Tumor MRI  (BTMRI)  \cite{cheng2015enhanced}. Our selection of these datasets is based on their diverse modalities and lesion characteristics, which allow us to test the broad applicability of our method across different medical imaging datasets. The Messidor dataset contains 1,200 retinal color fundus images with diabetic retinopathy four grading annotations. The ISIC2019 dataset consists of 25,331 gold-standard dermoscopic images that diagnose eight categories of benign and malignant skin lesions. The brain tumor dataset comprises 3,064 MRI T1-weighted contrast-enhanced slices of three types of brain tumors, obtained from different brain cross-sectional images of 233 patients.

\subsubsection*{Metrics}
In our experiments, the original image used to generate an adversarial watermark image is correctly classified. We evaluate the diagnosis AI models using classification Accuracy, which represents the ratio of correctly classified images to the total number of examples. A lower accuracy means the performance of the adversarial watermark is better. ROC-AUC (Receiver Operating Characteristic - Area Under the Curve) is used to measure a model's ability to discriminate between positive and negative samples. The ROC-AUC value represents the area under the ROC curve and ranges from 0 to 1. $AUC=0.5$ is equal to random guessing. A higher value indicates better model performance in distinguishing between positive and negative samples. We adopt micro-average OvR (One-vs-Rest) to calculate performance metrics in multi-class classification problems. To evaluate the image quality after watermark removal, some metrics are introduced: Peak Signal-to-Noise Ratio  (PSNR)  and Root Mean Square Error  (RMSE)  as indicators of image quality. Higher PSNR values indicate superior image quality, while lower RMSE values suggest higher quality. Structural Similarity  (SSIM)  is used to measure the similarity between two images. The range of SSIM is from 0 to 1, with higher values indicating greater similarity.

\subsubsection*{AI diagnostic models' training}
We employ a variety of backbone network architectures for the source models, covering Convolutional Neural Network  (CNN)  and Transformer models, such as ResNet50 \cite{he2016deep}, VGG16 \cite{simonyan2014very}, Inceptionv3 \cite{szegedy2015going}, MobileNetv3 \cite{howard2017mobilenets}, Densenet121 \cite{huang2018densenet}, and ViT \cite{dosovitskiy2021image}. For the Messidor and Brain Tumor datasets, we partition the data into training/validation sets using a 3:1 ratio and train the models with cross-entropy loss. In the case of ISIC2019, due to category imbalance, direct training would lead to the model assigning higher weights to the majority classes and lower accuracy for minority classes. Therefore, we resampled the dataset and trained the models using Focal Loss\cite{lin2017focal} instead of cross-entropy.

\subsubsection*{Lesion mask}
 In our experiments, we use the last convolutional layer to generate GradCAM-guided masks, ResNet50 for Messidor and ISIC2019, and DenseNet121 for BTMRI. In Figure~\ref{cam}, we exhibit the GradCAM heatmaps generated on three datasets. Following Algorithm \ref{algo2}, we threshold the GradCAM images to generate masks. As seen in Appendix~\ref{secA1}, Figure~\ref{figa1}, the accuracy drops as the threshold grows. A higher threshold leads to a more concentrated attention region. However, a lower threshold reduces the space of optimizable positions, which will affect the effectiveness of adversarial watermarks. We set threshold $t=190$.

\subsubsection*{Hyper-parameters of Watermark Optimization}
According to table~\ref{tab1}, HE-SIT shows the best performance, thus we use HE-SIT to optimize positions and transparencies for MIAD-MARK. Appendix~\ref{secA1} investigates how the hyper-parameters affect the performance of adversarial watermarks. The result shows that the watermark scale $sl$, position sampling rate $r_p$, population $N_p$, and crossover rate $r_c$ are the primary factors. We set $sl=2, N_p=15, N_g=2, s_{\alpha}=15, r_c=0.3, r_p=0.005$. 

\subsubsection*{Non-erasable Watermark Processing}
 Firstly, We fine-tuned two advanced DRNs  (SplitNet\cite{kim2017splitnet} and WDNet\cite{liu2021wdnet})  to ensure that they can remove watermark logos on medical images. These DRNs contain two-stage networks: a watermark region prediction network for locating logos and predicting compositions of alpha blending, and a refined network for watermark region recovery. Specifically, we train the decoders of DRNs with a composited medical image dataset where each image is blended with a logo, and the position and transparency are randomly sampled from uniform distributions. We separated this dataset into the training and the testing set with a 3:1 ratio. The training epoch is set to 50. The $\delta$ is generated in two ways according to the difference of the optimization objective, as Equation ~\ref{eq3} and Equation~\ref{eq4}. We employ the most effective settings for producing $\delta$, the complete experiments to obtain the best settings are displayed in Appendix~\ref{secB}. 
The perturbation range  $\delta$  is restricted by $L_{\infty}$ norm from 0 to 8/255. The step of  $\delta$  is set to 2/255, and each perturbation iterates 50 times.

\bibliography{sn-bibliography}

\newpage
\begin{appendices}

\section{Appendix}

\subsection*{Algorithm}\label{secA0}
In this section, we show pseudocodes of core algorithms, including Alpha blending \ref{algo1} for compositing medical images with watermark logos, the Generation of GradCAM-guided mask \ref{algo2}, and the heuristic evolutionary algorithm \ref{algo3} for watermark parameters optimization.
\begin{algorithm}
\caption{Alpha blending}\label{algo1}
\begin{algorithmic}[1]
\Require Original image ${I_{ori}}^{W\times H}$, logo ${I_{logo}}^{w\times h}$, perturbation $\delta$, position$ (x,y)$, transparency $\alpha$, logo scaling $sl$.\\
{Scaling ${I_{logo}}^{w\times h}$ by Equation~\ref{eq5} $\rightarrow$ ${I_{logo}}^{w^{'}\times h^{'}}$}.\\
{Create a zero background ${I_{b}}^{W\times H}$}.
\For {$i=x...x+w^{'}, j=y...y+h^{'}$}
\State
    ${I_{b}}^{W\times H} (i,j)  +{I_{logo}}^{w^{'}\times h^{'}} (i,j)  \rightarrow {I_{f}}^{W\times H} (i,j) $.
\EndFor \\
{Create a mask of ${I_{f}}^{W\times H} \rightarrow I_{m}^{W\times H} (i,j)  \in \{0,1\}$}.\\
{Get watermark image by Equation~\ref{eq6}: ${I_{w}}=\alpha/255 \cdot ( {I_{f}} - {I_m}\cdot  (I_{ori}+\delta) )  +  (I_{ori}+\delta) $}.\\
\Return $I_{w}$.
\end{algorithmic}
\end{algorithm}

\begin{algorithm}
\caption{GradCAM-guided mask generation}\label{algo2}
\begin{algorithmic}[1]
\Require Original image $I_{ori}$, ground truth label $l_g$, CNN model $C$, target convolution layer $L_{conv}$, sampling rate of channel $r$, threshold $t$.
\For {each $I_{ori}$}
\State
    Resize and normalize $I_{ori}$.
    \State Predicted class $l_p=argmax (C (I_{ori}) ) $.
     \If{$l_p=lg$}
        \State Gradient back-propagation according to Equation~\ref{eq7}.
        \State Average pooling to get weights $\Omega_n=\{\omega_1,...\omega_n\}$.
        \State Sampling features from $L_{conv}$ with $r\rightarrow \Lambda_k=\{A_0,...,A_k\},k\leq n$.
        \State Get Grad-CAM by Equation~\ref{eq8} 
        \State  Create a binary mask and resize  $\rightarrow {I_{M}}^{W\times H} (i,j) \in \{0,255\}$.
    \EndIf
\EndFor\\
\Return ${I^{0}_{M},...,I^{m}_{M}}$.
\end{algorithmic}
\end{algorithm}

\begin{algorithm}
\caption{Watermark parameters' optimization}\label{algo3}
\begin{algorithmic}[1]
\Require MIAD-MARKed image $I_w$, ground truth label $l_g$, GradCAM-guided mask $I_{M}$, Seen model $S$, Sample rate of position $r_p$, population $N_p$, maximum generation $N_g$, mutation step of alpha $s_{\alpha}$, crossover rate $r_c$.\\
{Get bounding boxes from $I_{M} \rightarrow B=\{b1,...,b_n\}$}.\\
$\mathcal{U}=\{ (x_0,y_0) ,..., (x_m,y_m) |0\leq x_m \leq W-w, 0\leq y_m\leq H-h\}$.
\Ensure $ (x,y)  \in \overline{B} \cap \mathcal{U}, \alpha \in  (0,255]$.
\For {each $I_w$}
    \State $Uniform (0,1] \rightarrow \alpha$.
    \State Calculate positions by Equation~\ref{eq9} $\rightarrow  (x_n,y_m) $.
    
    \State Calculate mutation step of position by Equation~\ref{eq10}  $\rightarrow s_p=[s_p (x) ,s_p (y) ]$.
    \State $\rho= (x_n,y_m,\alpha) ,
    P=\rho_0,\rho_1,...,\rho_q ,q<N$.
    \For {each individual $\rho_q$ in $P$}
        \State Algorithm\ref{algo1} $\rightarrow I_{w}$.
        \State $S (I,l_g)  \rightarrow f_t (I_w) ,f_t (I_{w}) ,l_p$.
        \If {$l_p \neq l_g$}
            \State
            \textbf{Break}
        \ElsIf {$f_t (I_{w}) <f_t (I_w) $}
            \State Select $N_p$ dominants $P^{'}=\rho_0,\rho_1,...,\rho_{N_p}$.
        \EndIf
    \EndFor

    \State$i=1,criterion=f_t (I_w) _{min}$.
    
    \While {$l_p = l_g$ and $i\leq N_g$}
        \State Mutate as Equation5 $\rightarrow U={\rho_0,\rho_1,...\rho_h,h<[len (\rho) ]^3\cdot N_p}$.
        \State Crossover $\rightarrow V={\rho_0,\rho_1,...\rho_k},k<h\cdot r_c$.
        \For {each individual $\rho_k$ in $V$}
        \State Algorithm\ref{algo1} $\rightarrow I_{w}^{k}$.
        \State $S (I_w,l_g)  \rightarrow l_p,f_t (I_{w}^{k}) $.
            \If {$l_p \neq l_g$}
                 \State  
                 \textbf{Break}
            \ElsIf {$f_t (I_{w}^{k}) <criterion$}
            \State Select $N_p$ dominant $V^{'}=\rho_0,\rho_1,...,\rho_{N_p}$.
            \EndIf
        \EndFor
        \State $i+1,criterion=f_t (I_w^k) _{min}$.
    \EndWhile\\
    \Return $ (x,y) , \alpha$.
\EndFor
\end{algorithmic}
\end{algorithm}

\subsection*{Hyperparameters adjustment}\label{secA1}

\begin{figure}[!t]
  \centering

  \begin{subfigure}{0.8\textwidth}
    \includegraphics[width=1\textwidth]{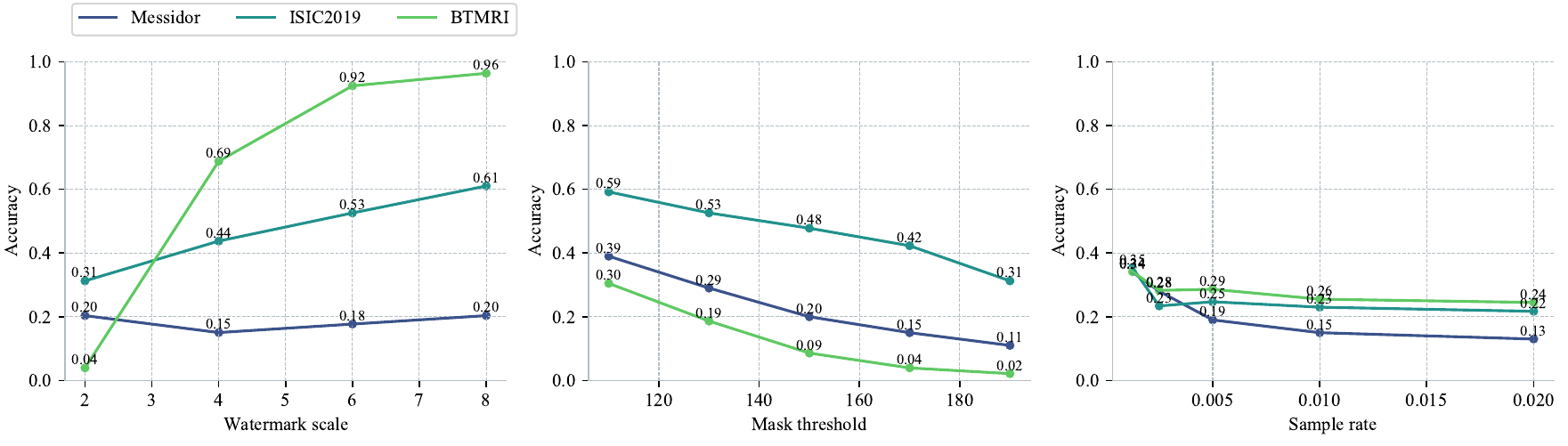} 
    \caption{Change watermark scale $sl$, mask threshold $t$, and sample rate $r_p$ to observe the impact on Accuracy.}\label{figa1}
  \end{subfigure}
  \hfill
  
  \begin{subfigure}{0.8\textwidth}
    \includegraphics[width=1\textwidth]{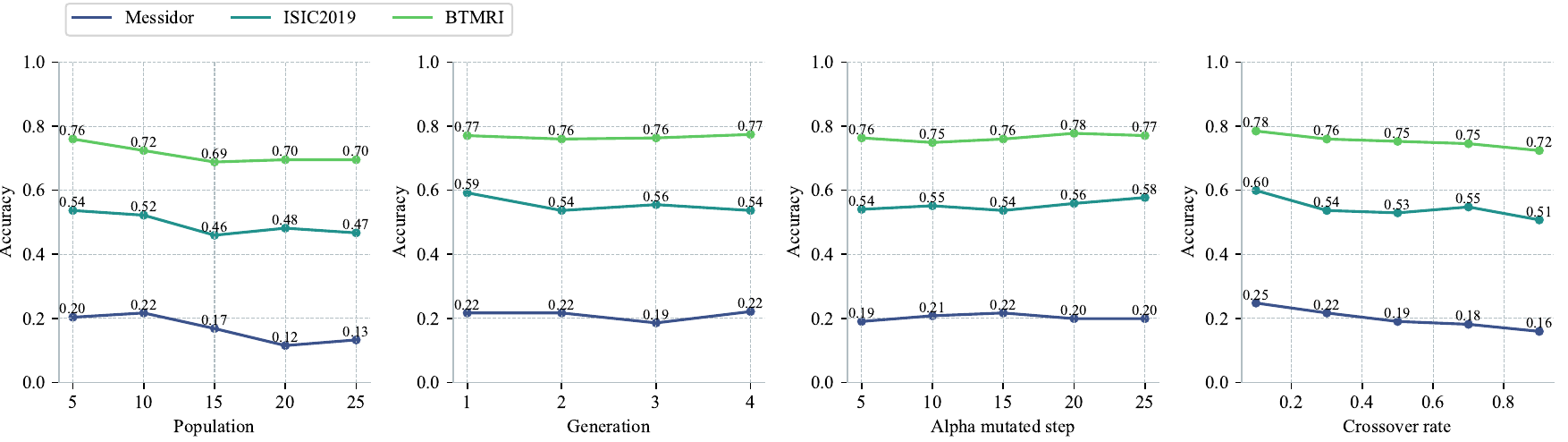} 
    \caption{Change population $N_p$, generation $N_g$, alpha mutated step $s_\alpha$, and crossover rate $r_c$ to observe the impact on Accuracy. The four hyperparameters are related to our evolutionary algorithm.}\label{figa2}
  \end{subfigure}
  \hfill
    \caption{Hyperparameters adjustment of MIAD-MARK.}
    \label{hyper}
\end{figure}

To investigate the effect of hyperparameter settings on the watermark adversarial production, we randomly sampled 300 data from each dataset based on categorical distribution. Figure~\ref{figa1} shows the most important three hyper-parameters that influence the classification accuracy. The watermark scale controls the watermark size, the accuracy is descent when the watermark size becomes larger. Messidor is insensitive to watermark size change, but the Accuracy of BTMRI drops dramatically as the watermark size becomes larger. Accuracy drops as the threshold of the attention map increases, because the restricted lesion areas become more concentrated, expanding the range within which watermark positions can be optimized. The sample rate $r_p$ controls the density of candidate positions. Accuracy drops as $r_p$ increases but the rate of improvement gradually levels off. In~Figure\ref{figa1}, expect the parameters being adjusted, the common settings for other parameters are  (1) Messidor: model='ResNet50', $logo='ADCIS', sl=2, t=190, N_p=15, N_g=2, s_{\alpha}=15, r_c=0.3, r_p=0.005$;  (2) ISIC2019: model='ResNet50', $logo='IBM', sl=2, t=190, N_p=15, N_g=2, s_{\alpha}=15, r_c=0.3, r_p=0.005$;  (3) BTMRI: model='DenseNet121', $logo='figshare', sl=2, t=190, N_p=15, N_g=2, s_{\alpha}=15, r_c=0.3, r_p=0.005$. For reproducibility of the experiment, we fix the random seed to '2023'.

In order to investigate the impact of other evolutionary algorithm hyperparameters on Accuracy, such as population size $N_p$, number of generations $N_g$, mutation step size of transparency $s_{\alpha}$, and crossover rate $r_c$, we modified the general parameter settings of Figure~\ref{figa1} to provide a larger search space for the watermark. This was done to better observe the effects of these parameters on Accuracy. Figure~\ref{figa2} shows that accuracy drops with population and crossover rate increases. These experiments follow common parameter settings:  (1) Messidor: model='ResNet50', $logo='ADCIS', sl=4, t=190, N_p=5, N_g=2, s_{\alpha}=15, r_c=0.3, r_p=0.001$;  (2) ISIC2019: model='ResNet50',$logo='IBM', sl=4, t=190, N_p=5, N_g=2, s_{\alpha}=15, r_c=0.3, r_p=0.001$;  (3) BTMRI: model='DenseNet121', $logo='figshare', sl=4, t=190, N_p=5, N_g=2, s_{\alpha}=15, r_c=0.3, r_p=0.001$. 

The approximate dynamic range of accuracy by tuning the hyper-parameters: the mask threshold is 30\%, the sample rate is 20\%, the population and crossover rate is 10\%, and the generation and transparent mutation step is 5\%.

\subsection*{Logos' impact}\label{Appendix_logos}

In this section, we investigate the influence of different logo styles on the performance of MIAD-MARK. We selected seven distinctive logos ('Beihang', 'ADCIS', 'IBM', 'SGH', 'figshare', 'ASTAR', 'nature') with varying shapes and complexities, as illustrated in Figure~\ref{logos}. Following the standardized configuration outlined in Appendix~\ref{secA1}, Figure~\ref{figa1}.

From the observations presented in Figure~\ref{logos}(a), it is evident that the average accuracy rates are as follows: 13.49\% for Messidor, 20.95\% for ISIC2019, and 45.44\% for BTMRI. The corresponding standard deviations (Std) are 7.25\% (Messidor), 8.99\% (ISIC2019), and 9.61\% (BTMRI). These results underscore the effectiveness of MIAD-MARK across a range of diverse logos. Moreover, the average inference numbers of models are 34 (Messidor), 44 (ISIC2019), and 128 (BTMRI), with corresponding Std values are 20 (Messidor), 34 (ISIC2019), and 30 (BTMRI). This indicates that the computational consumption remains within an acceptable range.

\begin{figure}[t]
\centering
\includegraphics[width=0.8\textwidth]{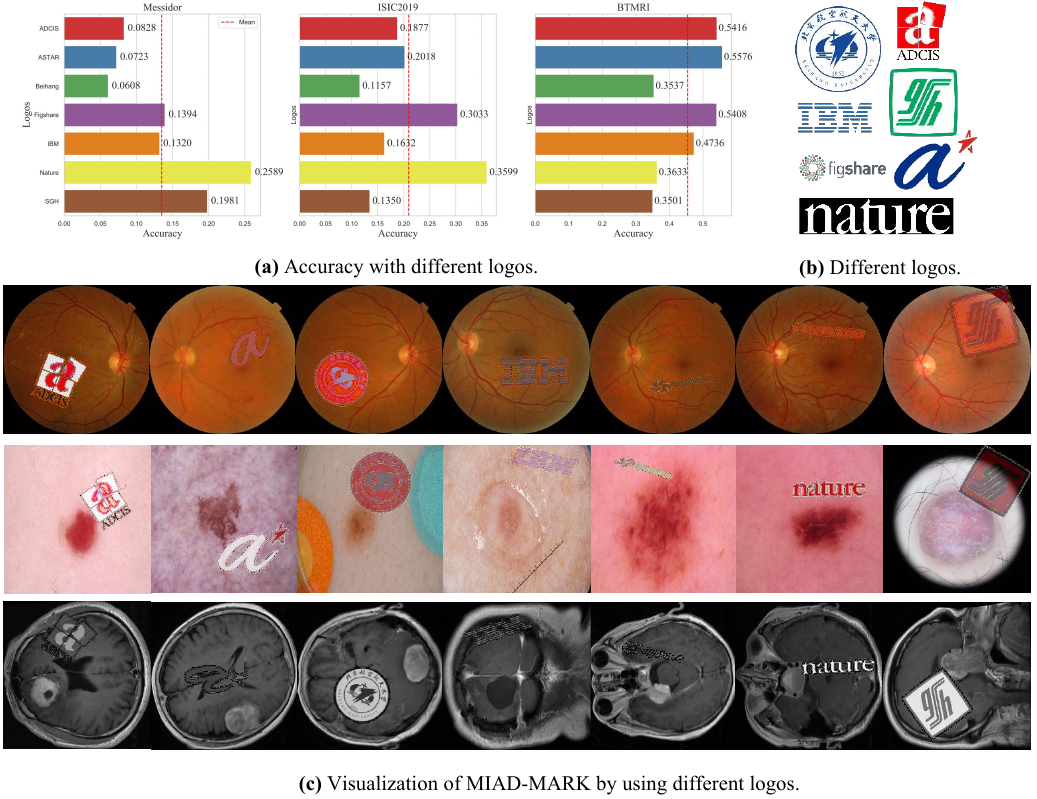} \caption{MIAD-MARK by using different logos. (a) AI diagnosis model accuracy with different logos on three datasets. The red dotted lines show the mean values. (b) Watermark logos used on figure(a) experiments. From left to right, top to bottom, the logo is: 'Beihang', 'ADCIS', 'IBM', 'SGH', 'figshare', 'ASTAR', 'nature'. (c) MIAD-MARKed medical images with different logos. The logos are transformed by SIT.}\label{logos}
\end{figure}

\subsection*{Discussion of  $\delta$  order settings}\label{secB}

\begin{figure}[htbp]
  \centering

  \begin{subfigure}{0.9\textwidth}
    \includegraphics[width=1\textwidth]{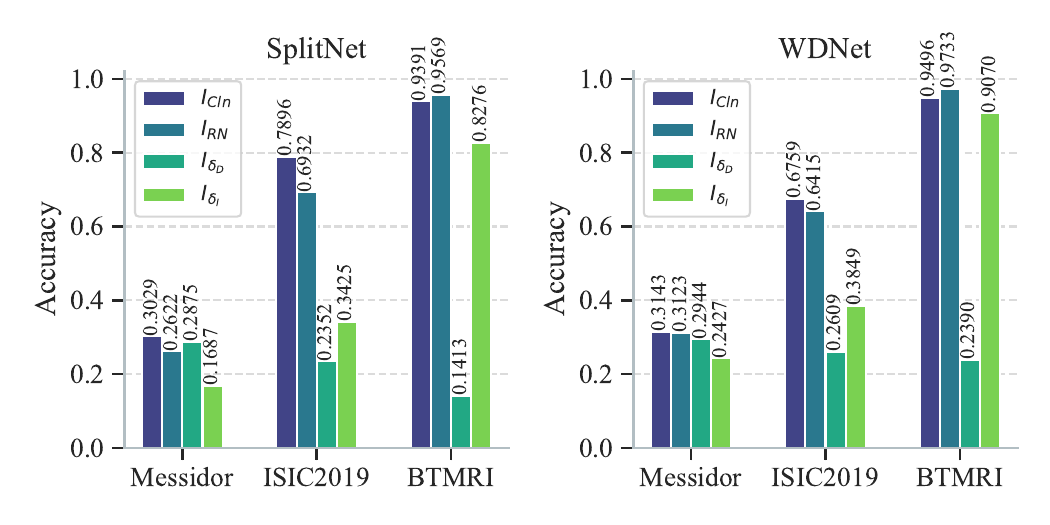} 
    \caption{Adding  $\delta$  before generating adversarial watermarks. This figure only shows the classification accuracy of DRNs' outputs, DRNs' inputs are adversarial watermarks and the accuracy equals 0.}\label{wv_first}
  \end{subfigure}
  \hfill
  
  \begin{subfigure}{1.0\textwidth}
        \centering
        \scalebox{0.8}{  
            \begin{tabular*}{\textwidth}{@{\extracolsep\fill}lccccccc}
            \toprule%
            && \multicolumn{2}{@{}c@{}}{Messidor} & \multicolumn{2}{@{}c@{}}{ISIC2019}  & \multicolumn{2}{@{}c@{}}{BTMRI}\\\cmidrule{3-4}\cmidrule{5-6}\cmidrule{7-8}%
             Perturbation setting &DRN & SplitNet & WDNet &  SplitNet & WDNet &  SplitNet & WDNet\\
            \midrule
            \multirow{2}*{$I_{Cln}$}& Input&0.0 &0.0 &0.0   &0.0  &0.0  &0.0  \\
            & Output&0.7665 &0.7116 &0.7938  &0.8321  &0.8281  & 1.0  \\
            \midrule
            \multirow{2}*{$I_{RN}$}& Input&0.2821 &0.2797 &0.3405  &0.3429  &0.0156  & 0.0156   \\
            & Output&0.4107 &0.5356 &0.6067  &0.6498  &0.0156  & 0.0260  \\
            \midrule
            \multirow{2}*{$I_{\delta_D}$}& Input&0.2934 &0.2747  &0.3381  &0.3429  &0.0156  & 0.0052   \\
            & Output&0.4531&0.4095  &0.2494  &0.1511  &0.0052 &0.0417  \\
            \midrule
            \multirow{2}*{$I_{\delta_I}$}& Input&0.2734 &0.2647 &0.3453  &0.3165  &0.0260  &0.0313    \\
            & Output&0.2772 &0.3283  &0.3429  & 0.3957 &0.0260  & 0.0417  \\
            \midrule
            \end{tabular*}}
        \caption{The classification accuracy of DRNs' inputs and outputs. Images are injected $\delta$ after generating adversarial watermarks.} \label{wv_after}
    \end{subfigure}
    
    \caption{Discussion about the order of adding $\delta$. $I_{Cln}$ means adding nothing in adversarial watermarking images. $I_{RN}$ means the image added random noise. $I_{\delta_D}$ represents the image added $\delta_D$ and the $I_{\delta_I}$ represents the images added $\delta_I$. $\delta_D$ is the perturbation of disrupting the images and $\delta_I$ is the  perturbation of inerasable watermark.}
    \label{wv_order}
\end{figure}

In this section, we will discuss the most effective perturbation settings for our adversarial watermarks on medical image datasets. '$I_{Cln}$' means adding nothing in adversarial watermarking images. '$I_{RN}$' means adding random noise to the image, which is used to control the noise variable, as $\delta$ is the optimized perturbation.

Firstly, we do experiments to decide the order of adding $\delta$ (Before or after adversarial watermark generation). When injecting $\delta$ after adversarial watermark generation, as seen in Table~\ref{wv_after}, none inputs are correctly classified in the '$I_{Cln}$' setting. But some input images are recognized in the '$I_{RN}$', '$I_{\delta_D}$', and '$I_{\delta_I}$' settings. This indicates that the injected noises invalidate some of the adversarial watermarks. Based on the above observation, we change the order to adding $\delta$ before adversarial watermark generation. Thus the adversarial watermark will not be influenced by $\delta$. As Seen in Figure~\ref{wv_first}, all inputs of DRN are predicted to the wrong classes and the figure shows the accuracy of the outputs. Adversarial watermark injected $\delta_D$ get the mean lowest accuracy, which indicates that $\delta_D$ effectively protects adversarial watermarks from deep removal. $\delta_I$ performs well on ISIC2019 but is less effective on the other two datasets. All settings on the Messidor dataset show a low accuracy. As seen in Appendix~\ref{secC}, the 'ADCIS' logo is more robust than 'IBM' and 'Figshare', because 'ADCIS' is more complicated after SIT, making it hard to remove clearly. Specifically, 'ADCIS' has text, graphics, and a variety of colors, but 'IBM' is a text logo and displays only one color. 'Figshare' is turned into the gray channel for grayscale brain tumor images, thus it can not implement hue variation, and the removal network can easily erase 'Figshare' and recover the accuracy of the classifiers. Overall, '$\delta_D$' is a more effective approach for watermark protection especially for the vulnerable logo whose appearance is simple and colorless.

Why is $\delta_I$ less effective than $\delta_D$? In Equation~\ref{eq3}, $\delta_I$ requires minimizing the distance between the predicted mask and the zero mask, which leads to no prediction of the watermark region. The aforementioned setup first adds vaccines and then generates adversarial watermarks. The watermark is randomly composited which differs from the actual watermark's pattern and position. This results in the $\delta$ being unable to prevent the removal of the network's predicted watermark mask. In conclusion, Generating unerasable adversarial watermarks on $\delta_D$-protected medical images is a more effective way to prevent deep removal, and does not weaken the threat of adversarial watermarks for DNN models.


\subsection*{Visualization after DRN processing}\label{secC}
Our experiment follows the common settings in the Appendix.~\ref{secA1}, Figure.~\ref{figa1}. We visualize the inputs, outputs, and predicted masks during the watermark removal process by SplitNet. As seen in Figure~\ref{clean}, medical adversarial watermarks without protection can be easily removed. Figure~\ref{rn} indicates that only random noise without optimization can not prevent deep removal. Figure~\ref{dwv} shows that $\delta_D$ can disrupt the outputs of the removal network, making the image unavailable. Figure~\ref{iwv} shows $\delta_I$ maintains the completion of watermarks.

Table~\ref{tab2} quantifies the visual quality impact of the $\delta$ on the dataset, including $\delta_I$ which does not appear in the main text. The results of random noise are not significantly different from those of the clean image, indicating that random noise does not affect the performance of the watermark removal network. The PSNR and SSIM scores of $\delta_D$ are the lowest, and the RMSE scores are the highest, indicating $\delta_D$ can severely degrade image quality after the watermark removal process. The PSNR and SSIM scores of $\delta_I$ are highest, and the RMSE score is lowest, indicating that the images processed by $\delta_I$ remain highly similar to the original image even after passing through the watermark removal network. The $RMSE$ of $\delta_I$ is significantly lower than the other settings, indicating that its watermark region is more likely to be preserved.

\begin{table}[ht]
\caption{Quantitative evaluation of visual quality and structural similarity in different perturbation settings. $I_{Cln}$ means adding nothing in adversarial watermarking images. $I_{RN}$ means images added random noise. $I_{\delta_D}$ represents images added $\delta_D$ and the $I_{\delta_I}$ represents images added $\delta_I$. $\delta_D$ is the perturbation of disrupting the images and $\delta_I$ is the  perturbation of inerasable watermark.}\label{tab2}
\begin{tabular*}{\textwidth}{@{\extracolsep\fill}lccccccc}
\toprule%
&& \multicolumn{2}{@{}c@{}}{Messidor} & \multicolumn{2}{@{}c@{}}{ISIC2019}  & \multicolumn{2}{@{}c@{}}{BTMRI}\\\cmidrule{3-4}\cmidrule{5-6}\cmidrule{7-8}%
 Metric& Perturbation setting& SplitNet & WDNet &  SplitNet & WDNet &  SplitNet & WDNet\\
\midrule
\multirow{4}*{$PSNR$}& $I_{Cln}$ &40.2835&	40.6051&	41.1074&	42.0666&	43.779&	45.0266
 \\
&$I_{RN}$  &40.8846&	40.5668&	41.0293&	43.5864&	43.8055&	32.0738

\\
&$I_{\delta_D}$ &38.7061&	32.0738&	36.9566&	34.0239&	37.4647&	30.6235

\\
&$I_{\delta_I}$  &45.1667&	43.5864&	43.7348 &	45.3694 & 44.3292&	46.3083

\\
\midrule
\multirow{4}*{$SSIM$}&$I_{Cln}$  &0.9334&	0.9421&	0.9508&	0.9537&	0.9625&	0.9689

\\
&$I_{RN}$ &0.9466&	0.9432&	0.9550&	0.9627&	0.9645&	0.6809

\\
&$I_{\delta_D}$ &0.9011&	0.6809&	0.8834&	0.8013&	0.8638&	0.5763
\\
&$I_{\delta_I}$  &0.9813&	0.9627& 0.9825&	0.9732&	0.9738&	0.9740
\\ 
\midrule
\multirow{4}*{$RMSE$}& $I_{Cln}$ &2.4863&	2.4086	&2.2607	&2.0540	&1.6528	&1.4443

\\
&$I_{RN}$ &2.3362	&2.4184	&2.282	&1.8741	&1.6527	&6.6703

\\
&$I_{\delta_D}$&3.2013	&6.6703	&3.8327	&5.4753	&3.6003	&7.5914
\\
&$I_{\delta_I}$&1.5433	&1.8741	&1.7668	&1.6020	&1.5715	&1.2880
\\ 

\midrule
\end{tabular*}
\end{table}

\subsection*{Visualization of GradCAM-guided lesion mask}
\begin{figure}[!t]
\centering
\includegraphics[width=0.95\textwidth]{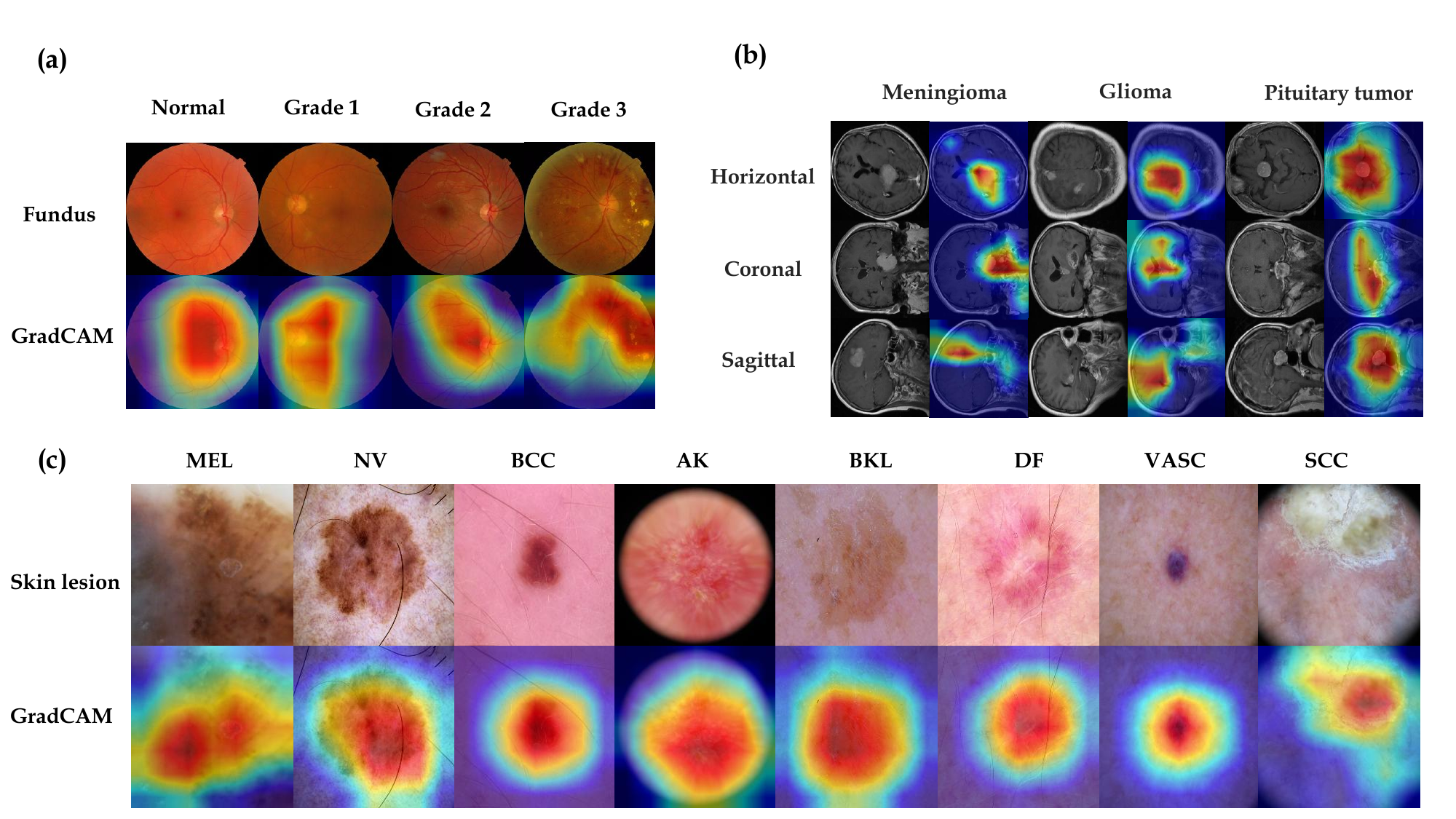} \caption{GradCAM heatmaps of three datasets. (a) Diabetic retinopathy grading heatmaps, the medical diagnosis is based on microaneurysms, hemorrhages, neovascularization, and no neovascularization. (b) Brain tumor MRI heatmaps, including three lesion types: Meningioma, Glioma, and Pituitary tumor. Each class has three MRI scan planes: Horizontal, Coronal, and Sagittal. (c) Eight-class skin lesion heatmaps.}\label{cam}
\end{figure}

\begin{figure}[!t]
  \centering

  \begin{subfigure}{0.75\textwidth}
    \includegraphics[width=1\textwidth]{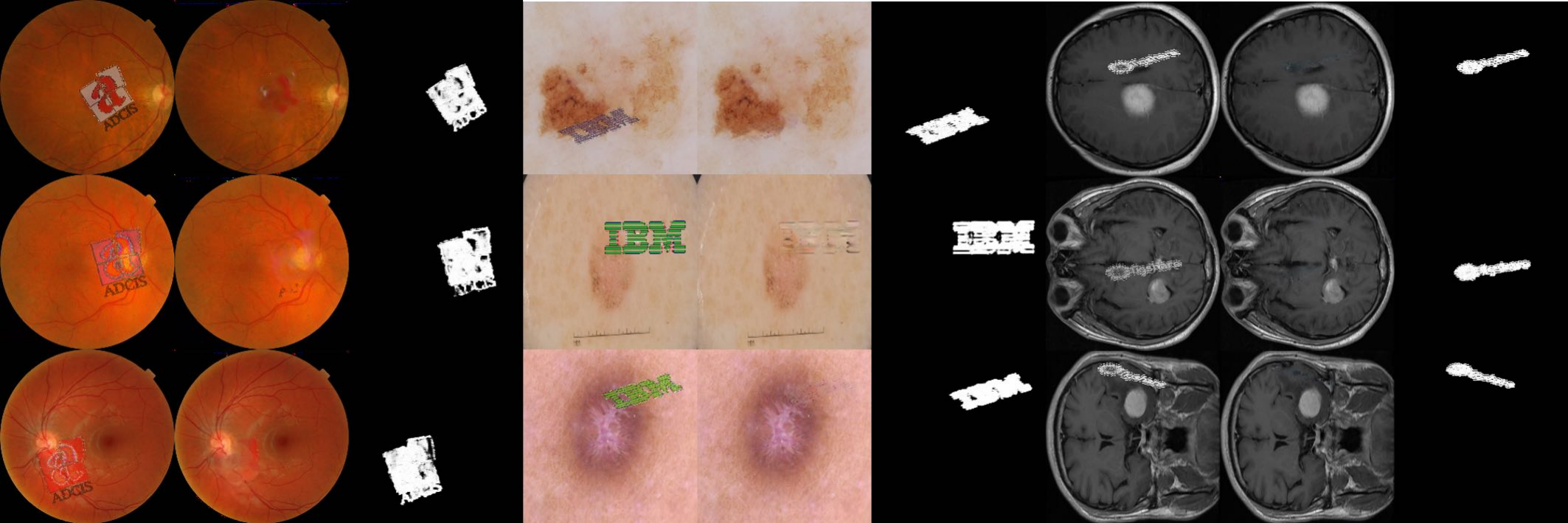} 
    \caption{Watermarked medical images without any protection. The watermark region is clearly displayed in predicted masks.}\label{clean}
  \end{subfigure}
  \hfill

  \begin{subfigure}{0.75\textwidth}
    \includegraphics[width=1\textwidth]{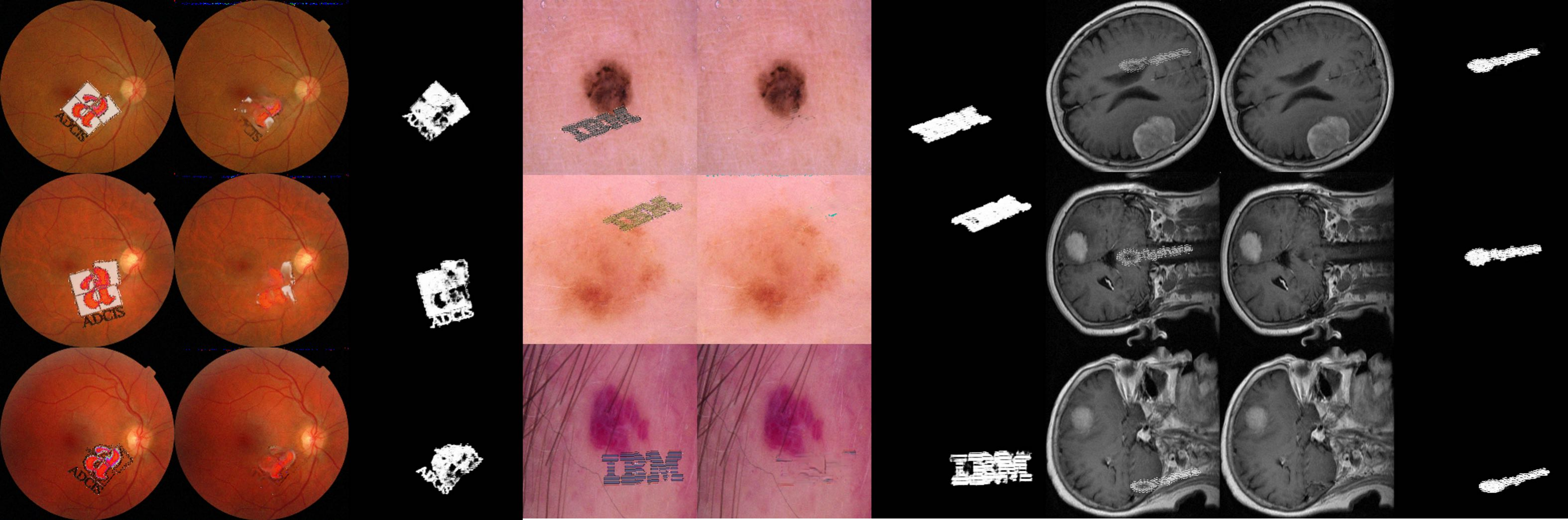} 
    \caption{Adversarial watermarking with random noise as a control experiment for $\delta$. Except for the fundus, watermarks in the other two datasets are almost completely removed.}\label{rn}
  \end{subfigure}
  \hfill

    \begin{subfigure}{0.75\textwidth}
    \includegraphics[width=1\textwidth]{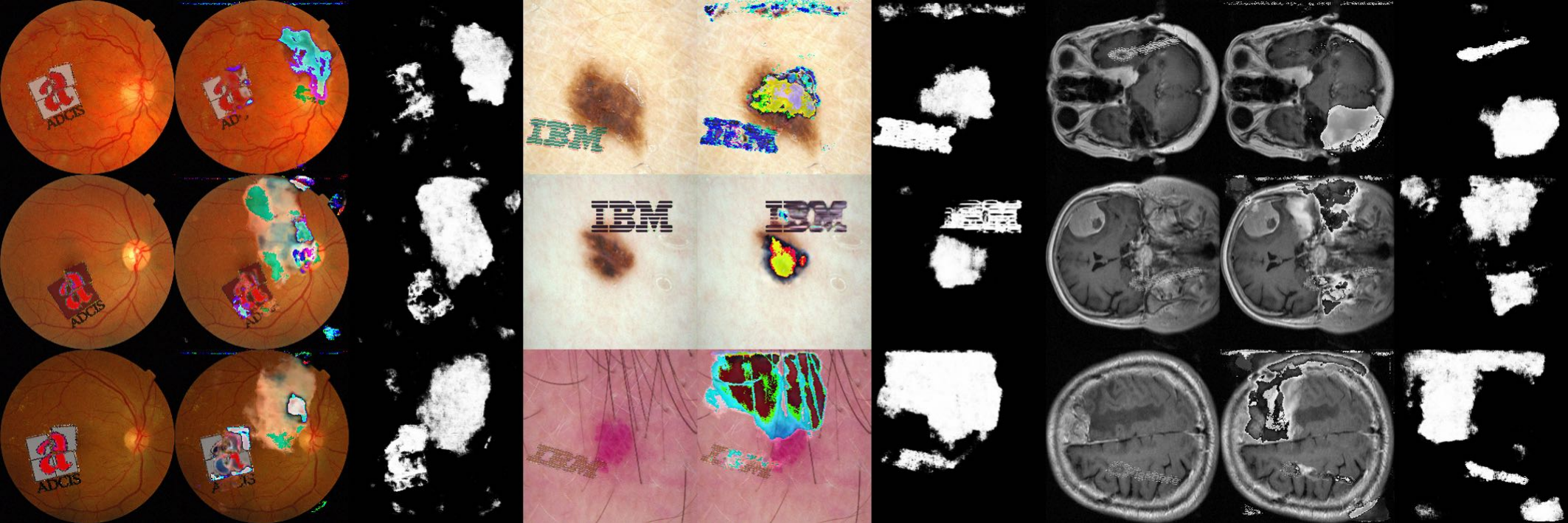} 
    \caption{$\delta_D$ leads to distortion of the output images. The predicted masks contain non-watermarked regions, the DRN processes these regions and makes corruption.}\label{dwv}
  \end{subfigure}
  \hfill

    \begin{subfigure}{0.75\textwidth}
    \includegraphics[width=1\textwidth]{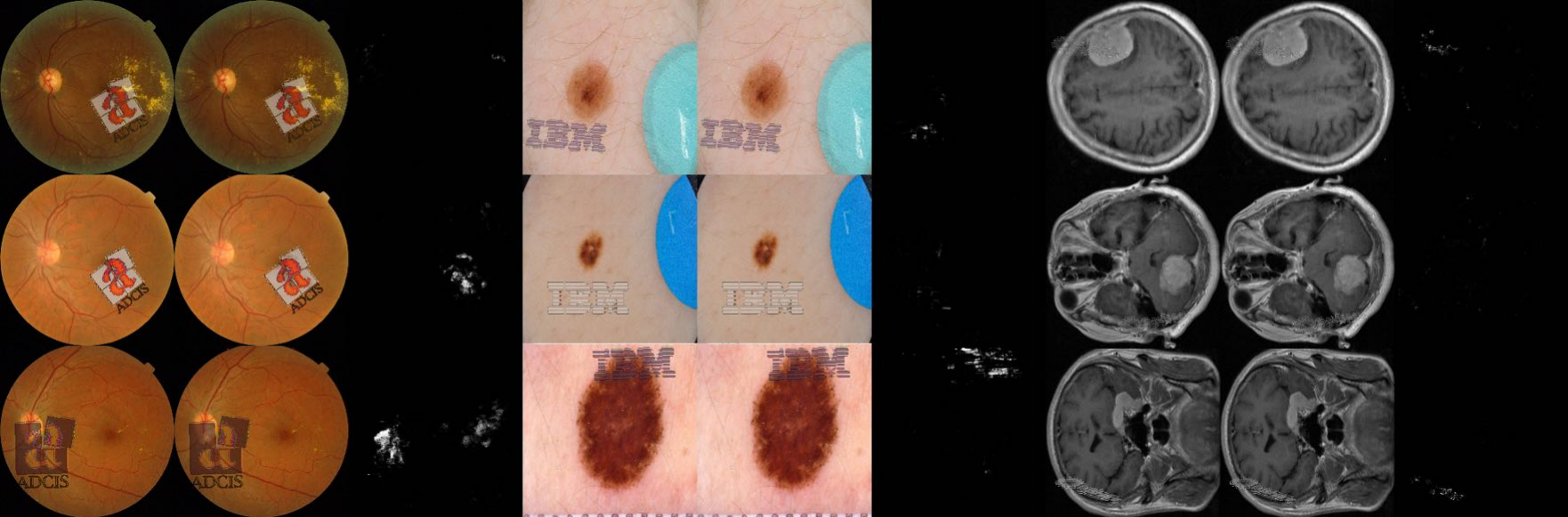} 
    \caption{$\delta_I$ maintains the watermarks. The predicted masks are nearly zero, which indicates that watermarks are not located correctly.}\label{iwv}
  \end{subfigure}

    \caption{Visualization of adversarial watermarks in different perturbation settings after DRN processing.}
    \label{drn_visual}
\end{figure}

In this part, we visualize GradCAM of three datasets to show its ability to locate lesion regions on medical images. As seen in Figure~\ref{cam}, subfigure (a) shows the GradCAM heatmap of the diabetic retinopathy grading on the Messidor dataset. Subfigure (b)  shows heatmaps of three classes of brain tumors. Subfigure (c) shows heatmap of eight classes on the ISIC2019 dataset.

\end{appendices}

\end{document}